\documentclass{article}

    \PassOptionsToPackage{round}{natbib}

\usepackage[preprint]{neurips_2024}

\usepackage[utf8]{inputenc} 
\usepackage[T1]{fontenc}    
\usepackage{hyperref}       
\usepackage{url}            
\usepackage{caption}
\usepackage{subcaption}
\usepackage{booktabs}       
\usepackage{amsfonts}       
\usepackage{nicefrac}       
\usepackage{microtype}      
\usepackage{xcolor}         

\usepackage{fancyhdr}
\usepackage{algorithm}
\usepackage{algorithmic}
\usepackage{natbib}
\usepackage{eso-pic} 
\usepackage{forloop}
\usepackage{wrapfig}

\usepackage{amsmath}
\usepackage{amssymb}
\usepackage{mathtools}
\usepackage{amsthm}

\usepackage{amsmath,amsfonts,bm}

\newcommand{\Dec}{A}

\newcommand{\X}{X} 

\newcommand{\loss}{\ell}
\newcommand{\Loss}{L}

\newcommand{\Sampling}{S}

\newcommand{\Unobserved}{U}
\newcommand{\smallprob}{\alpha}

\newcommand{\weight}{w}
\newcommand{\Weight}{W}

\newcommand{\policy}{\pi}
\newcommand{\odds}{\text{odds}}

\newcommand{\Prob}{\mathbb{P}}
\newcommand{\prob}{p}
\newcommand{\probhat}{\widehat{p}}
\newcommand{\probpolicy}{p_{\policy}}
\newcommand{\Probpolicy}{\mathbb{P}_{\policy}}
\newcommand{\oddshat}{\widehat{\odds}}

\newcommand{\setdata}{\mathcal{D}}
\newcommand{\settrial}{\mathcal{D}}

\newcommand{\setloss}{\mathcal{D}''}
\newcommand{\setweight}{\mathcal{D}'}

\newcommand{\usetwithtest}{\mathcal{S}_{+}}

\newcommand{\n}{n}

\newcommand{\nweight}{m'}

\newcommand{\ind}[1]{\mathds{1}(#1)}

\newcommand{\rct}{\textsc{rct}}
\newcommand{\xgboost}{\textsc{xgb}oost}

\def\eqref#1{equation~\ref{#1}}

\def\ceil#1{\lceil #1 \rceil}

\def\1{\bm{1}}

\DeclareMathAlphabet{\mathsfit}{\encodingdefault}{\sfdefault}{m}{sl}
\SetMathAlphabet{\mathsfit}{bold}{\encodingdefault}{\sfdefault}{bx}{n}

\newcommand{\E}{\mathbb{E}}

\mathtoolsset{showonlyrefs}
\usepackage{dsfont}
\usepackage{tikz}
\usetikzlibrary{shapes.geometric}
\usetikzlibrary{positioning}

\newtheorem{theorem}{Theorem}[section]

\usepackage[textsize=tiny]{todonotes}

\defcitealias{nhanes13}{NHANES}

\author{%
  Sofia Ek \\
  Uppsala University\\
  \texttt{sofia.ek@it.uu.se} \\
  \And
  Dave Zachariah \\
  Uppsala University\\
  \texttt{dave.zachariah@it.uu.se} \\
}

\title{Externally Valid Policy Evaluation from Randomized Trials Using Additional Observational Data}

\begin{document}

\maketitle

\begin{abstract}
Randomized trials are widely considered as the gold standard for evaluating the effects of decision policies. Trial data is, however, drawn from a population which may differ from the intended target population and this raises a problem of external validity (aka. 
generalizability). In this paper we seek to use trial data to draw valid inferences about the outcome of a policy on the target population. Additional covariate data from the target population is used to model the sampling of individuals in the trial study. We develop a method that yields certifiably valid trial-based policy evaluations under any specified range of model miscalibrations. The method is nonparametric and the validity is assured even with finite samples. The certified policy evaluations are illustrated using both simulated and real data.
\end{abstract}

\section{Introduction}
Randomized controlled trials (\rct) are often considered to be the `gold standard' when evaluating the effects of different decisions or, more generally, decision policies.  \rct{} studies circumvent the need to identify and model potential confounding variables that arise in observational studies. They enable the evaluation and learning of decision policies for use in, e.g., clinical decision support, precision medicine and recommendation systems \citep{qian2011performance,zhao2012estimating,kosorok2019precision}. 

However, \rct{}s sample individuals that may differ systematically from a target population of interest. For instance, clinical trials usually involve only individuals who do not have any relevant comorbidities and those who volunteer for trials may very well exhibit different characteristics than the target population. Invalid inferences about a decision policy can be potentially harmful in safety-critical applications, where the cautionary principle of ``above all, do no harm'' applies \citep{smith2005origin}. This is especially challenging since the distributions of population characteristics are unknown. How can we \emph{generalize} results from the trial sample to the intended population?

\begin{figure*}
    \centering
    \begin{subfigure}[b]{0.46\textwidth}
        \centering
        \includegraphics[width=1\linewidth]{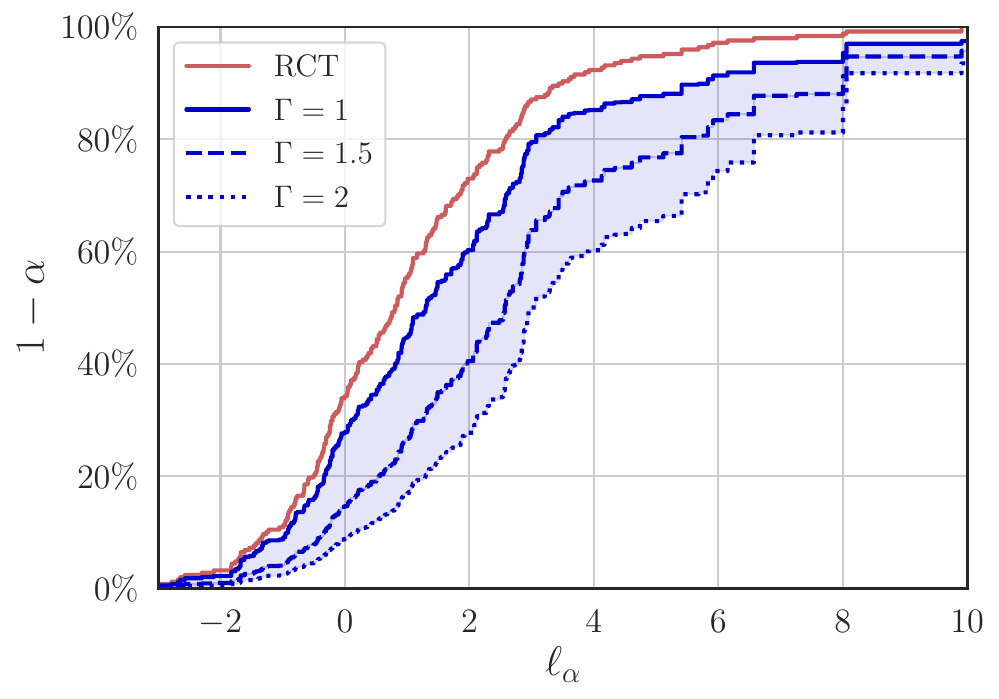} 
        \caption{}
         \label{fig:enter-label_loss}
    \end{subfigure}
    \hfill
    \begin{subfigure}[b]{0.46\textwidth}
        \centering
        \includegraphics[width=1\textwidth]{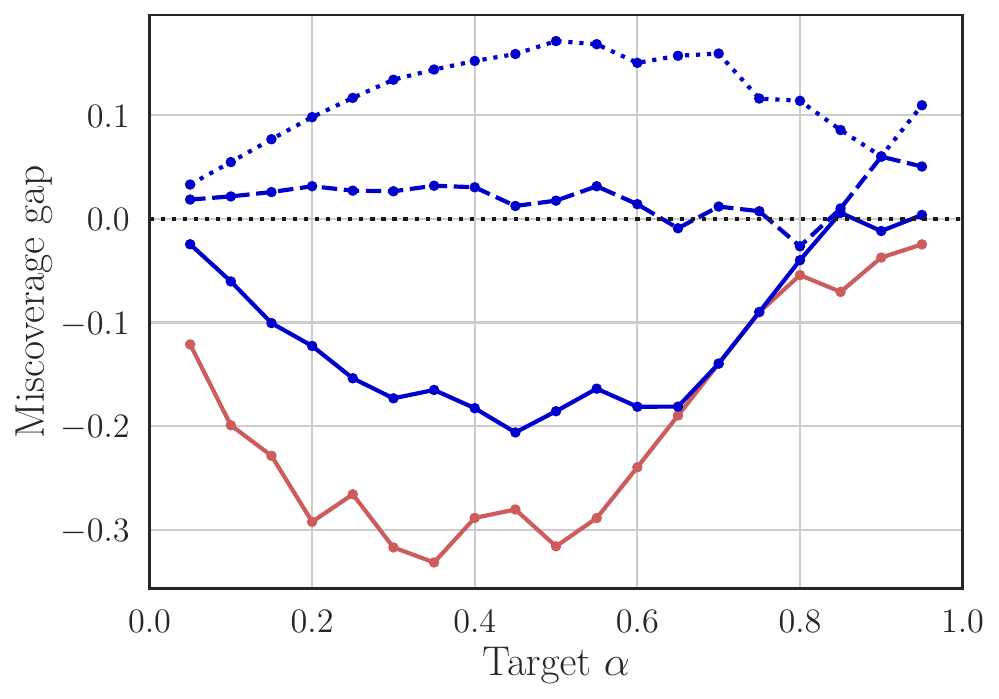}
        \caption{}
         \label{fig:enter-label_coverage}
    \end{subfigure}
    \caption{Inferring the out-of-sample losses of a policy $\policy$. (a) The loss $\Loss$ is bounded by an upper limit $\loss_{\alpha}$ with a probability of at least $1-\alpha$. The \rct{}-based limit curve uses only trial data, whereas the other limit curves also utilize a sampling model  trained using additional covariate data $\X$ from the target population. Each limit curve in blue is certified to provide valid inferences for models miscalibrated up to a degree $\Gamma$ defined in \eqref{eq:Gamma_divergence}. (b) Gap between the actual probability of exceeding the limit, $\Loss > \loss_{\alpha}$, and the nominal  probability of miscoverage $\alpha$. A negative gap means the inference $\loss_{\alpha}$ is \emph{invalid}, while a positive gap implies it is \emph{valid} but conservative. Details of the experiment are presented in Section~\ref{sec:synthetic-data}.}
    \label{fig:enter-label}
\end{figure*}

The focus of this paper is the problem of establishing \emph{externally} valid inferences about outcomes in a target population, when using experimental results from a trial population \citep{campbell1963experimental,manski2007identification,westreich2019epidemiology}. We consider evaluating a decision policy, denoted $\policy$, that maps covariates $\X$ of an individual onto a recommended action $\Dec$. The outcome of this decision has an associated loss $\Loss$ (aka. disutility or negative reward). Thus $\Loss$ may directly represent the outcome of interest. We assume the availability of samples $(\X, \Dec, \Loss)$ from the trial population and \emph{additional} covariate data $\X$ from the target population \citep{lesko2016effect,li2022generalizing,colnet2020causal}. The covariate data is used to model the sampling of individuals in the trial study. We propose a method for evaluating policy $\policy$ that
\begin{itemize}
    \item is nonparametric and makes no assumptions about the distributional forms of the data,

    \item takes into account possible covariate shifts from trial to target distribution, even when using miscalibrated sampling models with unmeasured selection factors,

    \item and certifies valid finite-sample inferences of the out-of-sample loss, up to any specified degree of model miscalibration.
\end{itemize}
Many policy evaluation methods are focused on estimating the expected loss $\E_{\policy}[\Loss]$ of $\policy$. However, since a substantial portion of losses $\Loss$ may exceed the mean, this focus can miss important tail events \citep{wang2018quantile,huang2021off}. By contrast, evaluating a policy in terms of its out-of-sample loss provides a more complete characterization of its performance and is consonant with the cautionary principle. Figure~\ref{fig:enter-label} illustrates the evaluation of $\policy$ using limit curves which upper bound the out-of-sample loss $\Loss$ with a given probability $1-\alpha$. A limit curve based on \rct{}-data alone is only ensured to be valid for a trial population. Using additional covariate data, however, we can certify the validity of the inferences for the target population up to any specified degree of miscalibration of the sampling model.

The rest of the paper is outlined as follows. We first state the problem of interest in Section~\ref{sec:problem} and relate it to the existing literature in Section~\ref{sec:background}. We then propose a policy evaluation method in Section~\ref{sec:method} and demonstrate its properties using both synthetic and real data in Section~\ref{sec:experiments}. We conclude the paper with a discussion about the properties of the method in Section~\ref{sec:discussion} and its broader impact in Section~\ref{sec:limitations}.

\section{Problem Formulation}
\label{sec:problem}
Any policy $\policy$, whether deterministic or randomized, can be described by a distribution $\probpolicy(\Dec|\X)$. Each covariate $\X$, unmeasured selection factor $U$ and action $\Dec$ has an associated loss $\Loss \in (-\infty, \Loss_{\max})$. We consider here a discrete action space, i.e. $\Dec \in \{1 \dots, K \}$. The decision process has a causal structure that can be formalized by a directed acyclic graph, visually summarized in Figure~\ref{fig:DAG_policy} \citep{peters2017elements}. The sampling indicator $\Sampling$ indicates whether individuals are drawn from a \emph{target} population, $\Sampling=0$, or a \emph{trial} population, $\Sampling=1$.  The causal structure allows us to decompose the two distributions. Specifically, the target distribution factorizes as
\begin{equation}
    \probpolicy(\X,\Unobserved, \Dec,\Loss|\Sampling=0) = \prob(\X, \Unobserved|\Sampling=0) \cdot \prob_{\policy}(\Dec|\X) \cdot \prob(\Loss|\X, \Unobserved, \Dec),
    \label{eq:jointdistribution_rwd}
\end{equation}
where only the policy $\probpolicy(\Dec|\X)$ is known. Similarly, the trial distribution factorizes as
\begin{equation}
    \prob(\X,\Unobserved, \Dec,\Loss|\Sampling=1) = \prob(\X, \Unobserved|\Sampling=1) \cdot \prob(\Dec|\X) \cdot \prob(\Loss|\X, \Unobserved, \Dec), 
    \label{eq:jointdistribution_rct}
\end{equation}
where only the randomization mechanism $\prob(\Dec|\X)$ is known. In general the characteristics of target and trial populations may \emph{differ}, that is, $p(\X, \Unobserved|\Sampling=0) \neq p(\X, \Unobserved|\Sampling=1)$. The unmeasured $\Unobserved$ may include self-selection factors that are challenging to record. Note, however, that only factors that also affect the loss $\Loss$ are relevant here.

\begin{figure}
\begin{subfigure}{0.45\linewidth}
\centering
        \begin{tikzpicture}[
        square/.style={regular polygon,regular polygon sides=4},
        observednode/.style={circle, draw=black!80, fill=white!5, very thick, minimum size=8.5mm},
        unobservednode/.style={circle, draw=black!80, dashed, fill=white!5, very thick, minimum size=8.5mm},
        fixednode/.style={square, draw=black!80, fill=white!5, very thick, inner sep = -0.12em},
        ]
        \node[fixednode]         (S)    {$\Sampling\!\!=\!\!0$};
        \node[observednode]      (X)    [below left = 0.2 cm and 0.7 cm of S]    {$\X$};
        \node[unobservednode]    (U)    [below right = 0.2 cm and 0.7 cm of S]   {$\Unobserved$};
        \node[observednode]      (A)    [below = 0.7 cm of X]    {$\Dec$};
        \node[observednode]      (L)    [below = 0.7 cm of U]   {$\Loss$};

        \draw[->, very thick, draw=black!80] (X) -- (A);
        \draw[->, very thick, draw=black!80] (A) -- (L);
        \draw[-, very thick, draw=black!80] (X) -- (U);
        \draw[->, very thick, draw=black!80] (U) -- (S.east);
        \draw[->, very thick, draw=black!80] (U) -- (L);
        \draw[->, very thick, draw=black!80] (X) -- (L);
        \draw[<-, very thick, draw=black!80] (S.west) -- (X);
        \end{tikzpicture}  
        \caption{}
\end{subfigure}
\hfill
\begin{subfigure}{0.45\linewidth}
\centering
        \begin{tikzpicture}[
        square/.style={regular polygon,regular polygon sides=4},
        observednode/.style={circle, draw=black!80, fill=white!5, very thick, minimum size=8.5mm},
        unobservednode/.style={circle, draw=black!80, dashed, fill=white!5, very thick, minimum size=8.5mm},
        fixednode/.style={square, draw=black!80, fill=white!5, very thick, inner sep = -0.12em},
        ]
        \node[fixednode]         (S)    {$\Sampling\!\!=\!\!1$};
        \node[observednode]      (X)    [below left = 0.2 cm and 0.7 cm of S]    {$\X$};
        \node[unobservednode]    (U)    [below right = 0.2 cm and 0.7 cm of S]   {$\Unobserved$};
        \node[observednode]      (A)    [below = 0.7 cm of X]    {$\Dec$};
        \node[observednode]      (L)    [below = 0.7 cm of U]   {$\Loss$};

        \draw[->, very thick, draw=black!80] (X) -- (A);
        \draw[->, very thick, draw=black!80] (A) -- (L);
        \draw[-, very thick, draw=black!80] (X) -- (U);
        \draw[->, very thick, draw=black!80] (U) -- (S.east);
        \draw[->, very thick, draw=black!80] (U) -- (L);
        \draw[->, very thick, draw=black!80] (X) -- (L);
        \draw[<-, very thick, draw=black!80] (S.west) -- (X);
        \end{tikzpicture}  
        \caption{}
\end{subfigure}
    \caption{Causal structure of process (a) under policy $\policy$  as well as (b) the trial study. Sampling indicator $\Sampling$ distinguishes between the two. For the important case of \rct{}, assignment of $\Dec$ is not influenced by any covariates so that the path $\X \rightarrow \Dec$ is eliminated.}
    \label{fig:DAG_policy}      
\end{figure}

From the trial distribution \eqref{eq:jointdistribution_rct}, we sample $m$ individuals $\settrial = \big((\X_i, \Dec_i, \Loss_i)\big)^{m}_{i=1}$
independently.  In addition, we also obtain $\n$ independent samples of \emph{covariate-only} data
$\big(\X_1, \X_2, \dots, \X_{\n}\big)$
from the target population \eqref{eq:jointdistribution_rwd}. Our aim is to infer the out-of-sample loss $\Loss_{n+1}$ for individual $n+1$ under any policy $\policy$. Specifically, we seek a loss limit $\loss_{\alpha}$ as a function of $1-\alpha$, such that $\Loss_{n+1}  \leq \loss_{\alpha}$ holds with probability $1-\alpha$ as illustrated by the `limit curves' in Figure~\ref{fig:enter-label_loss}.

The sampling pattern of individuals is described by $\prob(\Sampling|\X,\Unobserved)$. This distribution is unknown, but we assume that a model $\widehat{\prob}(\Sampling | \X)$ is available. This model was fitted using held-out data $\{(\X_j,\Sampling_j)\}$ employing either the conventional logistic model or any state-of-the-art machine learning models (as exemplified below). It can also be obtained from previous studies. There is, however, no guarantee that $\widehat{\prob}(\Sampling | \X)$ is  calibrated and it may indeed diverge from the unknown sampling pattern. Nevertheless, we want inferences about the out-of-sample loss to be valid also for miscalibrated models. We therefore express the degree of miscalibration in terms of the selection odds:
\begin{equation}
\frac{1}{\Gamma} \; \leq \; \underbrace{\frac{\prob(\Sampling = 0 | \X,\Unobserved)}{\prob(\Sampling = 1 | \X,\Unobserved)}}_{\text{unknown selection odds}} \bigg/ \underbrace{\frac{\widehat{\prob}(\Sampling = 0 | \X)}{\widehat{\prob}(\Sampling = 1 | \X)}}_{\text{nominal selection odds}} \; \leq \; \Gamma \qquad \text(a.s.)
\label{eq:Gamma_divergence}
\end{equation}
That is, the nominal selection odds can diverge by a factor $\Gamma$, where $\Gamma=1$ implies a perfectly calibrated model. This model includes all sources of errors (selection bias, model misspecification, estimation error). 

A limit $\loss^\Gamma_{\alpha}$ provides an \emph{externally valid} inference of $\Loss_{\n+1}$, up to any specified degree of miscalibration $\Gamma$, if it satisfies
\begin{equation}
\boxed{
    \Probpolicy \Big\{ \Loss_{\\n+1} \leq \loss^\Gamma_{\alpha}(\setdata) \: | \: \Sampling=0 \Big\} \geq 1- \smallprob, \quad \forall \alpha.}
\label{eq:certificate}
 \end{equation}
The problem we consider is to construct this externally valid limit $\loss^\Gamma_{\alpha}$. This limit allows us to infer the full loss distribution of a future individual with $\Loss_{n+1}$ under policy $\policy$, rather than merely the expected loss $\E_{\policy}[\Loss]$. Specifically, the tail losses are important in healthcare and other safety critical applications where erroneous inferences could be harmful, and a cautious approach when implementing new policies is needed. 

The limit curve $\loss^\Gamma_{\alpha}$ for policy $\policy$ is valid up to any declared degree of odds miscalibration $\Gamma$, which establishes the credibility of the policy evaluation, cf. \cite{manski2003identification}. While $\Gamma$ is unknowable, especially under unmeasured $\Unobserved$, we can use ideas from sensitivity analysis to guide its selection using measured data \citep{rosenbaum1983assessing, tan2006distributional} . Following the method in \cite{huang2024sensitivity}, we treat observed selection factors in $\X$ as unmeasured $\Unobserved$ in \eqref{eq:Gamma_divergence} to benchmark appropriate values for $\Gamma$, as detailed in subsection \ref{sec:benchmarking_gamma}.

By increasing the range of $\Gamma$, we certify the validity of the inference under increasingly credible assumptions on $\probhat(\Sampling|\X)$ \citep{manski2003identification}.  As the model credibility increases, however, the informativeness of the inferences decreases. Since the upper bound on the losses, $\Loss_{\max}$, is a trivial and uninformative limit, we may define the informativeness of $\loss^\Gamma_{\alpha}$ as
\begin{equation}
\text{Informativeness} = 1-  \inf\{ \smallprob : \loss_{\smallprob}^{\Gamma}(\setdata) < \Loss_{\max}  \}.
\label{eq:infomativeness}
\end{equation} That is, the right limit of a limit curve, which decreases with the degree of miscalibration. Figure~\ref{fig:enter-label_loss} shows curves that are valid for miscalibration in the range $\Gamma \in [1,2]$, where the informativeness is $95\%$ and $90\%$, respectively. The latter figure means that we can infer a non-trivial bound on the loss for $90\%$ of the target population.

\emph{Remark:} This paper addresses the evaluation of a given policy, whether proposed by a clinical expert or learned from historical data. By setting aside samples from an \rct{} study, a policy $\policy$ can be learned, and its out-of-sample performance evaluated using the proposed methodology.

\begin{figure}[tb]
    \centering
    \begin{subfigure}[b]{0.47\textwidth}
        \centering
        \includegraphics[width=1\linewidth]{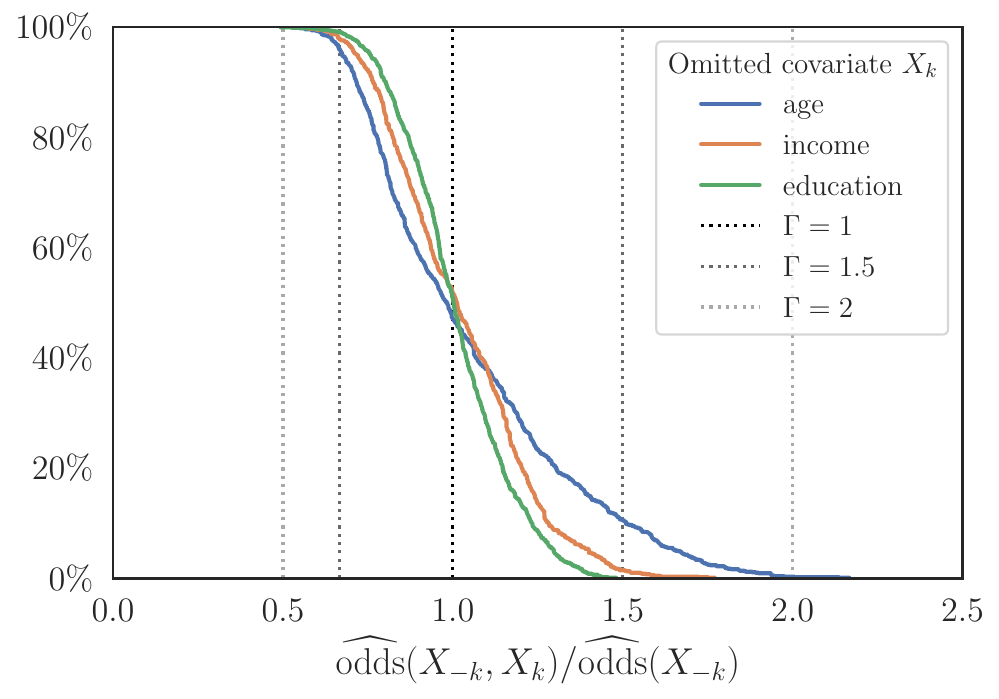}
        \caption{}
        \label{fig:nhanes_calibration}
    \end{subfigure}
    \hfill
    \begin{subfigure}[b]{0.47\textwidth}
        \centering
        \includegraphics[width=1\linewidth]{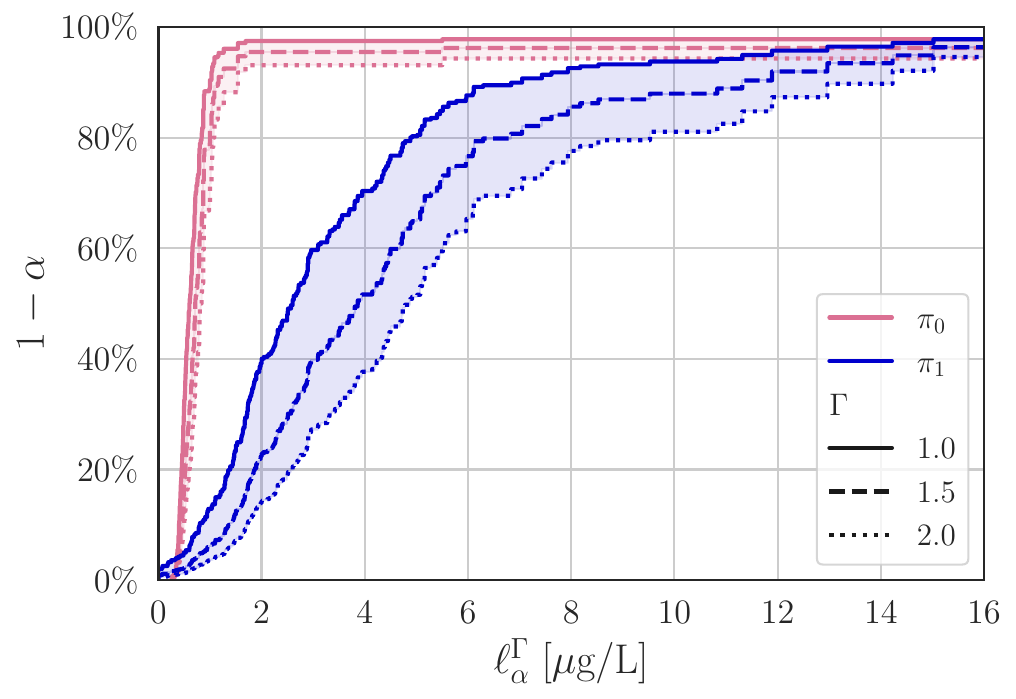}
        \caption{}
        \label{fig:nhanes_compare_policy}
    \end{subfigure}
    \caption{(a) Omitting measured selection factors to benchmark credible values for $\Gamma$ in \eqref{eq:Gamma_divergence}. (b) Inferred blood mercury levels [$\mu$g/L] in a target population under `high' and `low' seafood consumption ($\policy_1$ and $\policy_0$, respectively). Limit curves for degrees of odds miscalibration $\Gamma \in [1,2]$.}
    \label{fig:nhanes}
\end{figure}

\section{Background Literature}
\label{sec:background}
The problem considered herein is related to the problem of causal inference when combining data from randomized controlled trials and observational studies. Examples of the setting where only covariate data from the observational study is available can be found in \cite{lesko2016effect} and \cite{li2022generalizing}. For additional examples, refer to the survey by \cite{colnet2020causal}. 
Within the broader area of generalizability and transportability, the problem represents the case were the sampling probability depends solely on the covariates $\X$, and is independent of the action and the loss \citep{pearl2014external, lesko2017generalizing, degtiar2023review}. The problem is also related to a broader literature of statistical learning under covariate shifts, see for instance  \cite{shimodaira2000improving, sugiyama2007direct, quinonero2008dataset, reddi2015doubly, chen2016robust}.

The most common object of inference in policy evaluation is the expected loss $\E_{\policy}[\Loss]$ and a popular method for estimating it is inverse probability of sampling weighting (IPSW), which models covariate shifts from trial to target populations. The estimator using \rct{}-data is defined as
\begin{equation}
    V^{\policy}_{\text{IPSW}} = \frac{1}{n} \sum_{i = 1}^m  \frac{\probhat(\Sampling=0|\X_i)}{\probhat(\Sampling=1|\X_i)} \cdot \frac{\probpolicy(\Dec_i|\X_i)}{\prob(\Dec_i)} \cdot \Loss_i.
\label{eq:ipsw}
\end{equation}
This methodology has been applied in various studies, see for example \cite{cole2010generalizing, stuart2011use, westreich2017transportability, buchanan2018generalizing}. It is widely recognized that misspecified logistic models can introduce bias when estimating the weights \citep{colnet2020causal} and more recent works have suggested using flexible models, such as generalized boosted methods \citep{kern2016assessing}, instead. The counterpart to IPSW, used in off-policy evaluation with observational data, is inverse propensity weighting (IPW). The problem with misspecified logistic models also applies here when estimating the classification probabilities,  aka. propensity scores \citep{mccaffrey2004propensity, lee2010improving}. In this case, generalized boosted methods and covariate-balancing methods have shown to be promising alternatives \citep{setodji2017right, tu2019comparison}. The literature on IPSW and IPW is mainly focused on average treatment effect estimation, that is $\E_{\policy_1}[\Loss] - \E_{\policy_0}[\Loss]$ where $\policy_1$ and $\policy_0$ denote the `treat all' and `treat none' policies, respectively. By contrast, we want to certify the distributional properties of $\Loss_{\n+1}$  for any $\policy$, even under miscalibration of $\probhat(\Sampling|\X)$.

Conformal prediction is a distribution-free methodology focused on creating covariate-specific prediction regions that are valid for finite-samples \citep{vovk2005algorithmic, shafer2008tutorial}. The methodology was extended by \citet{tibshirani2019conformal} to also work for known covariate shifts. \citet{jin2021sensitivity} combined the marginal sensitivity methodology developed in \citet{tan2006distributional} with the conformal prediction for covariate shifts to perform sensitivity analysis of treatment effects in the case of unobserved confounding. Our methodology draws upon techniques in conformal prediction, but instead of providing covariate-specific prediction intervals under a policy $\policy$, we are concerned with evaluating \emph{any} $\policy$ over a target population.

When full identification of the causal effect is not possible due to unmeasured confounders, partial identification sensitivity analysis can be used to evaluate the robustness of the estimates. \cite{rosenbaum1983assessing} introduced a sensitivity parameter to bound the odds ratio of the probability of treatment. More recent work has extended this approach to account for treatment effect heterogeneity and non-binary treatments, as seen in \cite{tan2006distributional, shen2011sensitivity, zhao2019sensitivity, dorn2023sharp} among others. However, it is often challenging to interpret the absolute value of the sensitivity parameter in practical scenarios. 
To address this, recent research has proposed benchmarking, or calibrating, results by estimating the effects of unmeasured confounders, see for example \cite{hsu2013calibrating, franks2019flexible, veitch2020sense, de2024hidden}. Note that all these papers work with sensitivity analysis for observational studies. \cite{huang2024sensitivity} instead combines sensitivity analysis for generalization with benchmarking to determine reasonable values of $\Gamma$. 

A biased sample selection, similar to that described by \eqref{eq:Gamma_divergence}, was considered in the context of average treatment effect estimation by \cite{nie2021covariate}. In contrast to that work, our primary focus is on ensuring the validity of inferences regarding out-of-sample losses, even when dealing with finite training data. We achieve this using a sample-splitting technique.

\section{Method}
\label{sec:method}
Here we construct a limit $\loss^\Gamma_{\alpha}(\setdata)$ on the out-of-sample losses under policy $\policy$ that satisfies \eqref{eq:certificate} for any given specified degree of miscalibration $\Gamma$. 

\subsection{Benchmarking degree of miscalibration}
\label{sec:benchmarking_gamma}
The limit curve $\loss^\Gamma_{\alpha}(\setdata)$  holds up to the specified degree of odds miscalibration $\Gamma$. Although $\Gamma$ is inherently unknown, sensitivity analysis and methods for assessing the calibration of models can guide its estimation using available data. 

We start with a benchmarking method that specifically accounts for the potential impact of unmeasured selection factors, $\Unobserved$. Building on the approach in \cite{huang2024sensitivity}, we treat observed selection factors in $\X$  as proxies for unmeasured $\Unobserved$ in equation \eqref{eq:Gamma_divergence}, providing a benchmark for selecting suitable $\Gamma$ values. Specifically, let the the omitted selection factor $\X_k$ be $\Unobserved$ and $\X_{-k}$ denotes the remaining covariates. Then the ratio in \eqref{eq:Gamma_divergence} is estimated by dividing $\oddshat(\X_{-k}, \X_k) = \frac{\widehat{\prob}(\Sampling = 0 | \X)}{\widehat{\prob}(\Sampling = 1 | \X)}$ by $\oddshat(\X_{-k}) = \frac{\widehat{\prob}(\Sampling = 0 | \X_{-k})}{\widehat{\prob}(\Sampling = 1 | \X_{-k})}$. Figure~\ref{fig:nhanes_calibration} summarizes the distribution of $\oddshat(\X_{-k}, \X_k)/\oddshat(\X_{-k})$ using a real data set, where the omitted $\X_k$ is either age, income or education. We see that the corresponding ratios all fall within $\Gamma = 2$. We can therefore conclude that if any unmeasured selection factor $\Unobserved$ is weaker than the omitted factors, it is credible to set $\Gamma$ in the range of $1.5$ to $2$. The corresponding loss curves are shown in Figure~\ref{fig:nhanes_compare_policy}, which illustrates the blood mercury levels in a target population under policies of `high' respective `low' seafood consumption that can credibly be extrapolated from trial data. More details are available in Section~\ref{sec:semi-real data}. Note that this benchmarking method serves as a guide for assessing the influence of unmeasured selection factors $\Unobserved$~\citep{cinelli2019sense}. 

Without unmeasured selection factors, methods for assessing the calibration of models $\probhat(\Sampling|\X)$ discussed in, e.g., \citet{murphy1977reliability,naeini2015obtaining,widmann2019calibration} can guide the specified lower bound of the range of miscalibration. The nominal selection odds in~\eqref{eq:Gamma_divergence}, i.e., $\widehat{\odds}(\X) = \frac{\widehat{\prob}(\Sampling = 0 | \X)}{\widehat{\prob}(\Sampling = 1 | \X)}$,
can be quantized into several bins and for each bin the unknown selection odds, i.e., $\odds(\X) = \frac{\prob(\Sampling = 0 | \X)}{\prob(\Sampling = 1 | \X)}$,
can be estimated by counting the samples from both the target and trial distributions present. In the case of a well-calibrated model, estimated unknown odds should match the quantized nominal odds for each bin. To assess calibration, this process can be iterated across multiple ranges within the dataset and visualized through a reliability diagram, as exemplified in Figure~\ref{fig:reliability_testB}. Using~\eqref{eq:Gamma_divergence}, we have that 
\begin{equation}
    \frac{1}{\Gamma} \cdot  \widehat{\odds}(\X) \; \leq \; \odds(\X) \; \leq \; \Gamma \cdot  \widehat{\odds}(\X).
\end{equation}
Take the expectation with respect to $\X$, conditional on the nominal odds in a specified interval (or bin) $I$, so that:
\begin{equation}
    \frac{1}{\Gamma} \cdot  \E [\widehat{\odds}(\X) \; | \; \widehat{\odds}(\X) \in I] \; \leq \; 
\E [\odds(\X) \; | \; \widehat{\odds}(\X) \in I] \; \leq \; 
\Gamma \cdot  \E [\widehat{\odds}(\X) \; | \; \widehat{\odds}(\X) \in I].
\end{equation}
The expected odds is then estimated for each bin $I$ by counting samples from the target and trial distributions.

\subsection{Inferring out-of-sample limit}
We will now construct the limit $\loss^\Gamma_{\alpha}(\setdata)$. For this we need to describe the distribution shift from trial to target distribution for all samples, including the $n+1$ sample. We begin by considering the true distribution shift expressed using the ratio
\begin{equation}
\frac{\probpolicy(\X,\Unobserved, \Dec,\Loss|\Sampling=0)}{\prob(\X,\Unobserved, \Dec,\Loss|\Sampling=1)}.
\label{eq:ratio}
\end{equation}
Inserting the factorizations \eqref{eq:jointdistribution_rwd} and \eqref{eq:jointdistribution_rct} into this ratio shows that specifying the distribution shift requires the unknown (conditional) covariate distribution $\prob(\X,\Unobserved|\Sampling)$. We can, however, bound  \eqref{eq:ratio} using the model of the sampling pattern, $\widehat{\prob}(\Sampling | \X)$, as follows:
\begin{equation}
c \cdot \underline{\Weight}^{\Gamma} \leq \frac{\probpolicy(\X, \Unobserved, \Dec,\Loss|\Sampling=0)}{\prob(\X, \Unobserved, \Dec,\Loss|\Sampling=1)} \leq c \cdot \overline{\Weight}^{\Gamma},
\label{eq:ratio_bound}
\end{equation}
where
\begin{equation}
\underline{\Weight}^{\Gamma} =
\frac{1}{\Gamma} \cdot
\frac{\widehat{\prob}(\Sampling = 0 | \X)}{\widehat{\prob}(\Sampling = 1 | \X)} \cdot
\frac{\prob_{\policy}(\Dec|\X)}{\prob(\Dec|\X)}, \qquad
\overline{\Weight}^{\Gamma} =
\Gamma \cdot
\frac{\widehat{\prob}(\Sampling = 0 | \X)}{\widehat{\prob}(\Sampling = 1 | \X)} \cdot
\frac{\prob_{\policy}(\Dec|\X)}{\prob(\Dec|\X)},
\label{eq:weights_lower_upper}
\end{equation}
and $c= \frac{\prob(\Sampling = 1)}{\prob(\Sampling = 0)}$ is a constant. To see this, we note that the ratio in
\eqref{eq:ratio} can be expressed as
\begin{equation}
c \cdot \frac{\prob(\Sampling=0|\X, \Unobserved)}{\prob(\Sampling=1|\X, \Unobserved)} \cdot \frac{\probpolicy(\Dec|\X)}{\prob(\Dec|\X)},
\end{equation}
using Bayes' rule. The bound \eqref{eq:ratio_bound} follows by applying \eqref{eq:Gamma_divergence}. We proceed to show that the factors \eqref{eq:weights_lower_upper} are sufficient to construct an externally valid limit $\loss^\Gamma_{\alpha}$ for odds divergences up to degree $\Gamma$, similar to \cite{ek2023off}.

To ensure finite-sample guarantees, the trial data is randomly divided into two sets, 
$\settrial =  \setweight \cup \setloss$, 
with respective samples sizes of 
$\nweight$ and $m - \nweight$. The set $\setweight$ is used to construct
\begin{equation}
\overline{\weight}_{\beta}^{\Gamma}(\setweight) = \begin{cases} 
\overline{\Weight}^{\Gamma}_{[\ceil{(\nweight+1)(1-\beta)}]}, & (\nweight+1)(1-\beta) \leq \nweight, \\
\infty, & \text{otherwise},
\end{cases}
\label{eq:weightupper_conformal}
\end{equation}
where $\overline{\Weight}^{\Gamma}_{[\cdot]}$ is the upper limit in \eqref{eq:weights_lower_upper} evaluated over $\setweight$ and ordered $\overline{\Weight}^{\Gamma}_{[1]} \leq \overline{\Weight}^{\Gamma}_{[2]} \leq \dots \leq \overline{\Weight}^{\Gamma}_{[\nweight]}$. We show that \eqref{eq:weightupper_conformal} upper bounds the ratio \eqref{eq:ratio} for a future sample with probability $1-\beta$ for any choice of $\beta \in (0,1)$. The set $\setloss$ is used to construct a stand-in for the unknown cumulative distribution function of the out-of-sample loss: 
\begin{align}
\widehat{F}(\loss; \setloss, w) =
\frac{\sum_{i \in \setloss} \underline{\Weight}_i^{\Gamma} \ind{\Loss_i \leq \loss}}
{\sum_{i \in \setloss}  \big[\underline{\Weight}_i^{\Gamma} \ind{\Loss_i \leq \loss} + \overline{\Weight}^{\Gamma}_i \ind{\Loss_i > \loss } \big] + w},
\label{eq:cdf_alpha}
\end{align}
where $w>0$ is a free variable representing the unknown out-of-sample weight $\overline{\Weight}^{\Gamma}_{n+1}$ for a future sample. Based on \eqref{eq:weightupper_conformal} and \eqref{eq:cdf_alpha}, define $\loss_{\alpha, \beta}^{\Gamma}$ as the quantile function
\begin{equation}
\loss_{\alpha, \beta}^{\Gamma} = \inf \left\{\loss : \widehat{F}(\loss; \setloss, \overline{\weight}_{\beta}^{\Gamma}(\setweight)) \geq  \frac{1-\smallprob}{1-\beta}  \right\}.
\label{eq:computequantile}
\end{equation}
This enables us to construct a valid limit on the future loss $\Loss_{n+1}$ for any miscoverage probability $\smallprob \in (0,1)$.

\begin{theorem}
\label{thm:main} For any odds miscalibration up to degree $\Gamma$,
\begin{equation}
\loss_{\alpha}^{\Gamma}(\setdata) = \min_{\beta : 0 < \beta  <  \alpha } \loss_{\alpha, \beta}^{\Gamma},
\label{eq:ell_alpha}
\end{equation} is an externally valid limit on the out-of-sample loss $\Loss_{\n+1}$ of policy $\policy$. That is, \eqref{eq:ell_alpha} is certified to satisfy \eqref{eq:certificate}. 
\end{theorem}

\begin{figure}
    \centering
    \begin{minipage}{0.47\textwidth}
        \centering
        \includegraphics[width=0.975\linewidth]{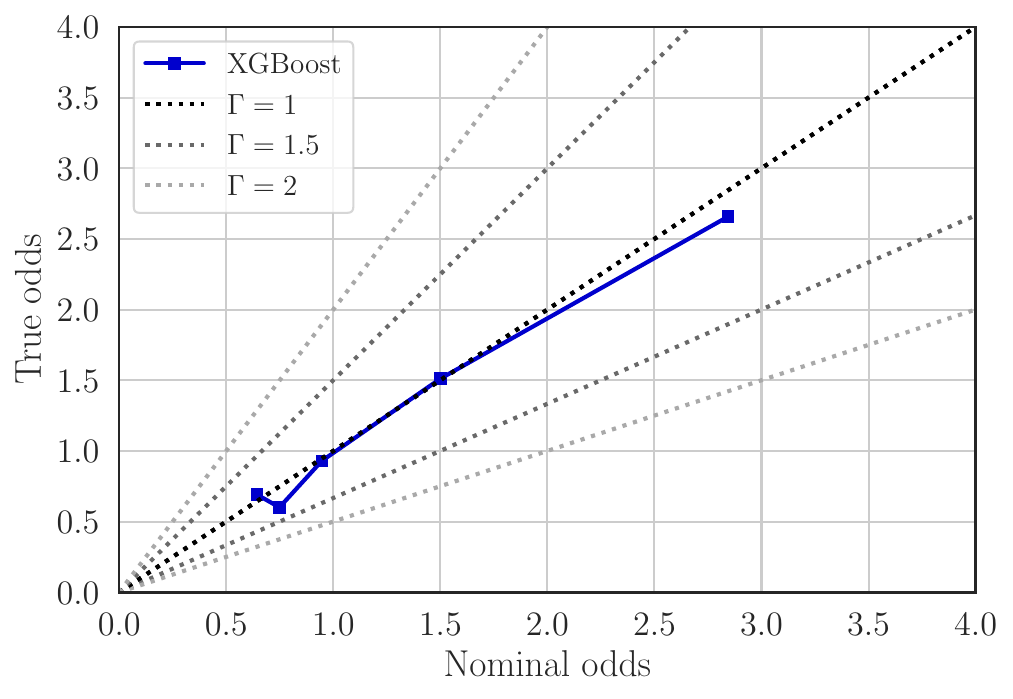}
        \caption{Reliability diagram of the observed odds against the average predicted nominal odds obtained from models $\probhat(\Sampling|\X)$.}
        \label{fig:reliability_testB}
    \end{minipage}%
    \hfill
    \begin{minipage}{0.47\textwidth}
        \centering
        \begin{algorithm}[H]
            \caption{A set of limit curves for policy $\policy$}     
            \label{alg:policy}
            \begin{algorithmic}[1]
                \INPUT Model $\probhat(\Sampling|\X)$, policy $\probpolicy(\Dec|\X)$, trial policy $\prob(\Dec|\X)$, a set of miscalibration degrees $\{1, \dots, \Gamma_{\text{max}}\}$ and trial data $\settrial$.
                \OUTPUT $\{ ( \Gamma, \smallprob, \loss_{\smallprob}^{\Gamma}) \}$
                \STATE Randomly split $\settrial$ into $\setweight$ and $\setloss$.
                \FOR {$\Gamma \in \{1, \dots, \Gamma_{\text{max}}\}$}
                    \FOR {$\smallprob \in \{0, \dots, 1 \}$}
                        \FOR {$\beta \in \{0, \dots, \smallprob \}$}
                        \STATE Compute $\overline{\weight}_{\beta}^{\Gamma}$ as in \eqref{eq:weightupper_conformal}.
                        \STATE Compute $\loss_{\alpha, \beta}^{\Gamma}$ as in \eqref{eq:computequantile}.
                        \ENDFOR
                    \STATE Set $\loss_{\alpha}^{\Gamma}$ to the smallest $\loss_{\alpha, \beta}^{\Gamma}$. 
                    \STATE Save $(\Gamma, \smallprob, \loss_{\smallprob}^{\Gamma})$.
                    \ENDFOR
                \ENDFOR
            \end{algorithmic}
        \end{algorithm}
    \end{minipage}
\end{figure}

The method seeks the level $\beta$ for the bound \eqref{eq:weightupper_conformal} that yields the tightest limit. The proof is presented in the supplementary material and builds on several techniques developed in \cite{vovk2005algorithmic, tibshirani2019conformal,jin2021sensitivity,ek2023off}.

Algorithm~\ref{alg:policy} summarizes the implementation  of a set of limit curves given a model of the sampling pattern $\probhat(\Sampling|\X)$ and a set of miscalibration degrees $[1,  \Gamma_{\max}]$. Note that  \eqref{eq:weightupper_conformal} and \eqref{eq:cdf_alpha} are step functions in $\beta$ respective $\ell$. Therefore \eqref{eq:computequantile} and \eqref{eq:ell_alpha} can be solved by computing the functions at a grid of points. Calculating \eqref{eq:weightupper_conformal} and \eqref{eq:cdf_alpha} requires the sorting of weights, but the sorting operation is a one-time requirement.

Increasing the degree of miscalibration $\Gamma$ results in a decrease in $\underline{\Weight}^{\Gamma}$  and an increase in  $\overline{\Weight}^{\Gamma}$ in  \eqref{eq:weights_lower_upper}. As weights associated with lower and higher losses decrease and increase, respectively, in \eqref{eq:cdf_alpha}, the resulting limit  $\loss_{\alpha}^{\Gamma}(\setdata)$ becomes more conservative.

\emph{Remark 1:} If the trial population is a small subgroup of the target population (for example in the case of weak overlap), the nominal odds $\frac{\widehat{\prob}(\Sampling = 0 | \X)}{\widehat{\prob}(\Sampling = 1 | \X)}$ will tend to be high. As this yields a significant weight $\overline{\weight}_{\beta}^{\Gamma}(\setweight)$, the informativeness \eqref{eq:infomativeness} of $\loss^\Gamma_{\alpha}(\setdata)$ diminishes. Nevertheless, the guarantee in \eqref{eq:certificate} holds.

\emph{Remark 2:} The split of $\settrial$ can be performed in a sample-efficient manner in the case of \rct{}, where actions are randomized so that $\prob(\Dec|\X) \equiv \prob(\Dec)$: For the $i$th sample, $(\X_i, \Dec_i, \Loss_i)$, draw $\widetilde{\Dec}_i \sim \probpolicy(\Dec|\X_i)$. Include sample $i$ in $\setloss$ if $\Dec_i = \widetilde{\Dec}_i$, otherwise include it in $\setweight$. This sample splitting method ensures that inferences on the loss are based on actions that match those of the policy.

\section{Numerical Experiments}
\label{sec:experiments}
We will use both synthetic and real-world data to illustrate the main concepts of policy evaluation with limit curves $(\smallprob,\loss^{\Gamma}_{\smallprob})$. As a performance benchmark, we estimate the quantile -- which yields the tightest limit -- using the inverse probability of sampling weighting \citep{colnet2020causal}
\begin{equation}
\loss_{\smallprob}(\setdata) = \inf \big\{ \loss :  \widehat{F}_{\text{IPSW}}(\loss; \setdata) \geq 1-\smallprob \big\},
\label{eq:quantile_empipw}
\end{equation}
where
\begin{align}
\widehat{F}_{\text{IPSW}}(\loss; \setdata) =
\frac{1}{m}  \sum_{i = 1}^m \frac{m}{n} \cdot \frac{\probhat(\Sampling\!=\!0|\X_i)}{\probhat(\Sampling\!=\!1|\X_i)} \cdot \frac{\probpolicy(\Dec_i|\X_i)}{\prob(\Dec_i)} \cdot \ind{\Loss_i \leq \loss}.
\label{eq:cdfest_ipsw}
\end{align}
This is similar to the approach in \citet{huang2021off} but adapted to problems involving data from trial and observational studies. 

We examine the impact of increasing the credibility of our model assumptions, i.e., by increasing the miscalibration degree $\Gamma$, on the informativeness \eqref{eq:infomativeness} of the limit curve. In addition, for the simulated data, we also assess the miscoverage gap of the curves
\begin{equation}
\text{Miscoverage gap} = \smallprob -  \Probpolicy\{ \Loss_{n+1} > \loss_{\smallprob}(\setdata)|\Sampling {=} 0\},
\label{eq:gap}
\end{equation}
where a negative gap indicates an invalid limit.

\subsection{Illustrations using synthetic data}
\label{sec:synthetic-data}
We consider target and trial populations of individuals with two-dimensional covariates, distributed as follows:
\begin{equation}
    \X | \Sampling = \begin{bmatrix} \X_{0, \Sampling} \\ \X_{1, \Sampling} \end{bmatrix} \sim \mathcal{N} \left( \begin{bmatrix}
        \mu_{0,\Sampling} \\
        \mu_{1,\Sampling}
    \end{bmatrix}, \begin{bmatrix}
        \sigma^2_{0, \Sampling} & 0 \\
        0 & \sigma^2_{1,\Sampling}
    \end{bmatrix} \right), \qquad 
    \Unobserved | \Sampling \sim \mathcal{N} (\mu_{\Unobserved,\Sampling}, \sigma^2_{\Unobserved,\Sampling}),
    \label{eq:sim_covariates}
\end{equation} 
where the parameters are given in Table~\ref{tab:simulations}. The distributions for populations $\texttt{A}$, $\texttt{B}$ and $\texttt{Trial}$ are taken to be unknown.

\begin{table}
    \caption{Means and variances of covariate distribution $\prob(\X, \Unobserved|\Sampling)$ in \eqref{eq:sim_covariates}.} 
    \medskip
    \centering
    \begin{tabular}{ccccccc}
        \toprule   
        Population & $\mu_{0, \Sampling}$ & $\mu_{1, \Sampling}$ & $\mu_{\Unobserved, \Sampling}$ & $\sigma_{0, \Sampling}^2$ & $\sigma_{1, \Sampling}^2$ & $\sigma_{\Unobserved, \Sampling}^2$ \\
        \midrule
        \texttt{A} $(\Sampling=0)$ & 0.5 & 0.5 & 0.5 & 1.0 & 1.0 & 1.0   \\ 
        \texttt{B} $(\Sampling=0)$ & 0.0 & 0.5 & 0.0 & 1.25& 1.5 & 1.25   \\  
        \midrule
        \texttt{Trial} $(\Sampling=1)$ & 0.0 & 0.0 & 0.0 & 1.0 & 1.0 & 1.0   \\
        \bottomrule
    \end{tabular}
    \label{tab:simulations}
\end{table}

The actions are binary $\Dec \in \{ 0, 1\}$ and corresponds to `do not treat' versus `treat'. We evaluate the `treat all' policy, i.e. $\prob_{\pi_1}(\Dec = 1|\X) = 1$. 
The trial is an \rct{} with equal probability of assignment, i.e., $\prob(\Dec)  \equiv 0.5$. The unknown conditional loss distribution is given by
\begin{align}
   \Loss|(\Dec, \X, \Unobserved) &\sim  \mathcal{N}(\Dec \cdot \X_0^2 + \X_1 + \Dec \cdot \Unobserved + (1-A), 1).
\end{align}
The sampling probability $\prob(\Sampling|\X, \Unobserved)$ is treated as unknown. The complete set of hyperparameters used is provided in the supplementary material. 

For the observational data $\big(\X_i\big)^{\n}_{i=1}$, we drew $n = 2000$ samples. For the trial data we drew $1000$ samples: $m=500$ samples and $\big((\X_i, \Dec_i, \Loss_i)\big)^{m}_{i=1}$ was used to compute the limit curves. The remaining $500$ samples were used to train $\widehat{\prob}(\Sampling|\X)$.

\begin{figure}[tb]
\centering
    \begin{subfigure}[b]{0.47\textwidth}
        \centering
         \includegraphics[width=1.0\linewidth]{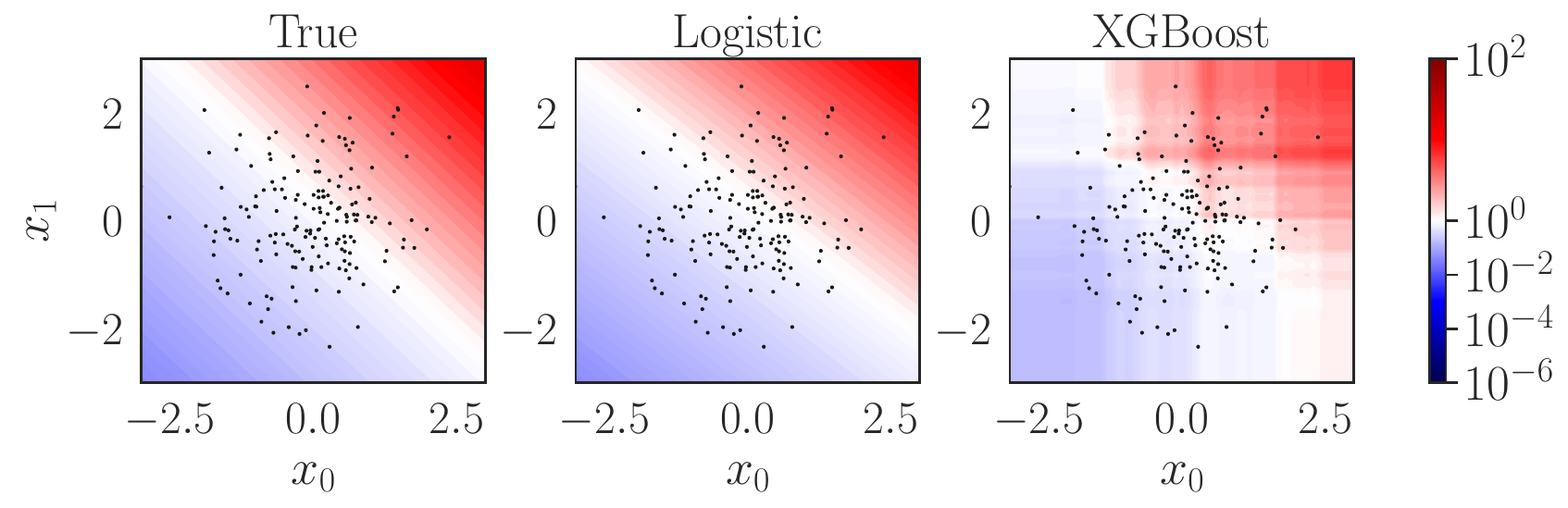}
        \caption{Selection odds for target population \texttt{A}.}
         \label{fig:synth_odds_test1}
    \end{subfigure}
    \hfill
    \begin{subfigure}[b]{0.47\textwidth}
        \centering
        \includegraphics[width=1.0\linewidth]{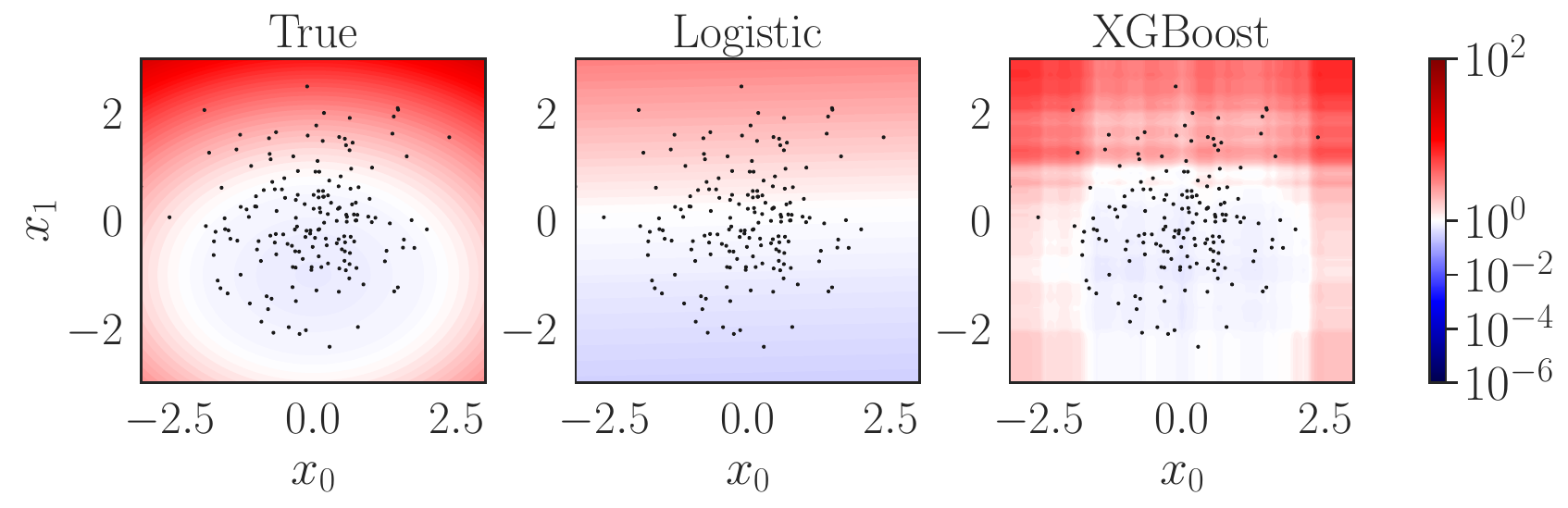}
        \caption{Selection odds for for target population \texttt{B}.}
        \label{fig:synth_odds_test2}
    \end{subfigure}
    \caption{Odds $\prob(\Sampling=0|\X)/\prob(\Sampling=1|\X)$ compared with nominal odds obtained from logistic and \xgboost{} models $\probhat(\Sampling|\X)$. The dots are a random subsample of the trial samples.}
\end{figure}

Figure~\ref{fig:synth_odds_test1} compares nominal selection odds obtained from the fitted models with the unknown odds $\prob(\Sampling=0|\X)/\prob(\Sampling=1|\X)$ for target population \texttt{A}. We consider two different fitted models $\probhat(\Sampling|\X)$: a logistic model, which is conventionally used in the causal inference literature \citep{westreich2017transportability}, and the more flexible tree-based ensemble model trained by \xgboost\ \citep{chen2016xgboost}. In this case, the logistic model happens to be well-specified so that the learned odds approximate the true ones well. The more flexible \xgboost\ model also provide visually similar odds, albeit less accurate. By contrast, Figure~\ref{fig:synth_odds_test2} repeats the same exercise for target population \texttt{B}. Here the logistic model is misspecified and severely miscalibrated while the \xgboost\ model continues to provide visually similar odds. A well-performing and flexible model is required for a meaningful benchmark of the upper value of $\Gamma$. Therefore, we will use the \xgboost\ model.

\begin{figure*}[tb]
    \centering
    \begin{subfigure}[b]{0.47\textwidth}
        \centering
        \includegraphics[width=1\linewidth]{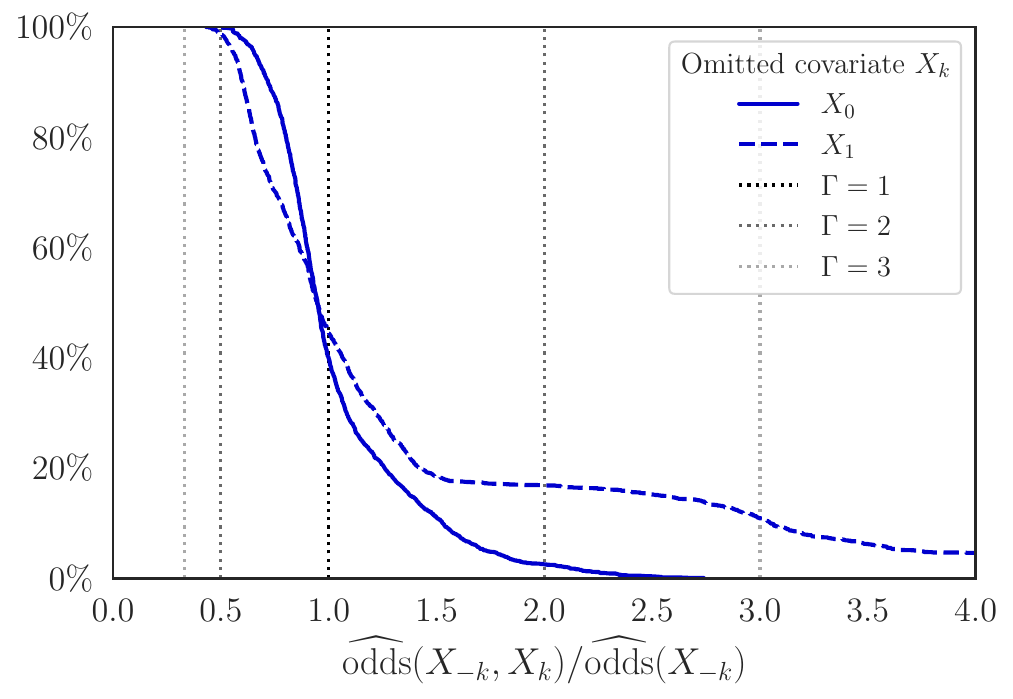} 
        \caption{}
        \label{fig:calibration_test2}
    \end{subfigure}
    \hfill
    \begin{subfigure}[b]{0.47\textwidth}
        \centering
        \includegraphics[width=1\textwidth]{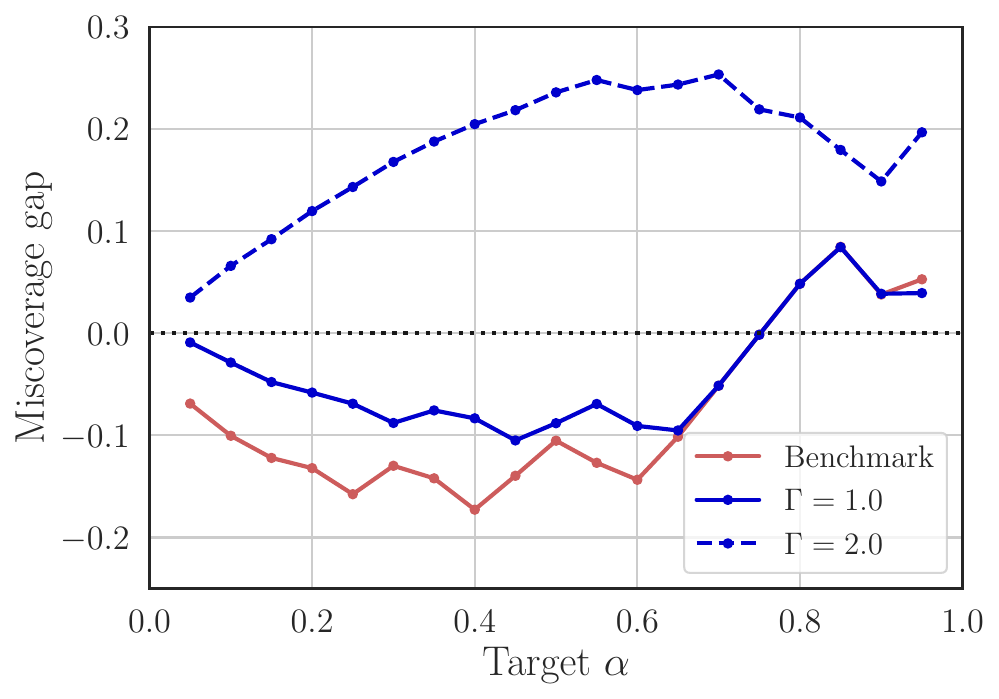}
        \caption{}
        \label{fig:synth_miscoverage_test2}
    \end{subfigure}
    \caption{Evaluating a `treat all'  policy $\policy_1$ for target population $\texttt{B}$. (a) Benchmarking the degree of miscalibration $\Gamma$ using omitted covariates. (b) Miscoverage gaps when degree of miscalibration $\Gamma \in  [1, 2]$.}
\end{figure*}

In Figure~\ref{fig:reliability_testB}, we use the reliability diagram technique to assess the performance of the \xgboost{} model $\widehat{\odds}(\X)$ of the nominal odds for target population \texttt{B}. The model is close to the diagonal suggesting that it is sufficiently flexible to accurately model the odds. This is in line with the result in Figure~\ref{fig:synth_odds_test2}. We then use observed covariates to calibrate appropriate upper bounds for $\Gamma$. Figure~\ref{fig:calibration_test2} shows the evaluation with respect to population \texttt{B}. Assuming that the unmeasured $\Unobserved$ have no greater selection effect than $\X_0$, a degree of miscalibration $\Gamma=2$ is a credible choice as it covers more than $95\%$ of the ratios. Since the data is simulated, we evaluate the miscoverage gap of the limit curves of the `treat all' policy $\policy_1$ in Figure~\ref{fig:synth_miscoverage_test2} for the benchmark and the limit curves for the proposed method. The gap is estimated using 1000 independent runs and for each run drawing $1000$ independent new samples $(\X_{n+1},\Unobserved_{n+1},\Dec_{n+1},\Loss_{n+1})$. We see that the benchmark and the limit curve for $\Gamma = 1$ yields a negative miscoverage gap. As the degree of miscalibration $\Gamma$ increases to 2, the limit curves exhibit positive miscoverage gaps.

The evaluation of $\policy_1$ with respect to population \texttt{A} is shown in Figure~\ref{fig:enter-label}. Results for the logistic model, another policy $\policy_0$, and for additional populations are available in the supplementary material.

\subsection{Evaluating seafood consumption policies}
\label{sec:semi-real data}
To illustrate the application of policy evaluation with real data, we study the impact of seafood consumption on blood mercury levels with data from the 2013-2014 National Health and Nutrition Examination Survey (NHANES). Following \citet{zhao2019sensitivity}, each individual's data includes eight covariates, encompassing gender, age, income, the presence or absence of income information, race, education, smoking history, and the number of cigarettes smoked in the last month. All covariates, except for smoking history $\Unobserved$, are treated as measured and denoted $\X$.  We excluded one individual due to missing education data and seven individuals with incomplete smoking data. We impute the median income for 175 individuals with no income information. The continuous covariates -- age, income and number of cigarettes smoked in the last month -- are standardized. After the preprocessing, our data set comprises 1107 individuals. The data is then split into observational data $\setdata_0$ and trial data $\setdata_1$ based on the covariates gender, age, income and smoking history (more details are available in the supplementary material) resulting in $646$ samples in the observational data and $461$ samples in the trial data. The action $\Dec$ describes individual fish or shellfish consumption, categorizing an individual as having either low ($\leq$1 serving in the past month) or high ($>$12 servings in the past month) consumption. The loss $\Loss$ represents the total concentration of blood mercury (measured in $\mu$g/L). 

To generate counterfactual actions and losses, we consider a balanced \rct{} so that $\prob(\Dec)  \equiv 0.5$ and  learn a model of $p(\Loss|\Dec, \X, \Unobserved)$ from the data using gradient boosting \citep{freund1997decision, friedman2000additive}. Thus during training, the trial data consists of samples $(\X_i,\Dec_i, \Loss_i)$ whereas the observational data only contains $\X$.

In Figure~\ref{fig:nhanes_calibration} we use observed covariates to benchmark appropriate upper bounds for $\Gamma$. We exclude each of the seven covariates one by one, highlighting the three most dominant ones in the figure. If the unmeasured covariates have weaker influence than these, a $\Gamma$ value in the range of $1.5$ to $2$ is appropriate. In Figure~\ref{fig:nhanes_compare_policy} we compare the limit curve for a policy $\policy_0$ corresponding to low $(\Dec\equiv 0)$ seafood consumption with the limit curve for a policy $\policy_1$ corresponding to high $(\Dec\equiv 1)$ seafood consumption. We use an \xgboost{}-trained model $\probhat(\Sampling|\X)$. For reference, a mercury level of 8 $\mu$g/L is guidance limit for women of child-bearing age. We see that under a low consumption policy a lower mercury level can be certified for miscalibrated odds $\Gamma \in [1,2]$. In this case, we can infer a 90\% probability that the out-of-sample mercury level falls below the reference value of 8 $\mu$g/L.

\section{Discussion}
\label{sec:discussion}
We have proposed a method for establishing externally valid policy evaluation based on experimental results from trial studies. The method is nonparametric, making no assumptions on the distributional forms of the data. Using additional covariate data from a target population, it takes into account possible covariate shifts between target and trial populations, and certifies valid finite-sample inferences of the out-of-sample loss $\Loss_{n+1}$, up to any specified degree of model miscalibration.

Conventional policy evaluation methods focus on $\E_{\policy}[\Loss]$ and can easily introduce a bias without the user's awareness, particularly when the model of the sampling pattern $\probhat(\Sampling|\X)$ is misspecified. Lacking any control for miscalibration undermines the possibility to establish external validity. In safety-critical applications, making invalid inferences about a decision policy can be potentially harmful. Hence, adhering to the cautionary principle of ``above all, do no harm'' is important. The proposed method is designed with this principle in mind, and the limit curve represents the worst-case scenario for the selected degree of miscalibration $\Gamma$. 

We also exemplify how the reliability diagram technique and the benchmark approach of omitting observed selection factors facilitate a more systematic guidance on the specification of the odds miscalibration degree $\Gamma$ in any given problem.

\section{Broader Impact and Limitations}
\label{sec:limitations}
The method we propose is designed for safety-critical applications, such as clinical decision support, with the cautionary principle in mind. We believe that it offers a valuable tool for policy evaluation in such scenarios. Our approach focuses on limit curves, coupled with a statistical guarantee, for a more detailed understanding of the out-of-sample loss. This facilitates a fairer evaluation by bringing attention to sensitive covariates in the tails. However, it remains important to be aware of biases, and it might be necessary to address them separately to prevent their replication. It is also important to note that the method requires independent samples, a condition that may not be met during major virus outbreaks or similar situations.


\begin{ack}
This work was partially supported by the Wallenberg AI, Autonomous Systems and
Software Program (WASP) funded by the Knut and Alice Wallenberg Foundation, and the Swedish Research Council under contract 2021-05022.
\end{ack}

\bibliography{sample}
\bibliographystyle{plainnat}

\newpage
\appendix
\section*{Supplementary Material}
In this supplementary material, we provide the proof outlined in Section~\ref{sec:proof} and additional details on the numerical experiments discussed in Section~\ref{sec:experiments_app}.

\section{Proof}
\label{sec:proof}

In this context, $\Prob$ represents the probability over samples drawn from both $\prob(\X, \Unobserved, \Dec,\Loss|\Sampling=1)$ and $\probpolicy(\X, \Unobserved, \Dec,\Loss|\Sampling=0)$. The proof technique presented here builds upon several results established in \cite{vovk2005algorithmic,tibshirani2019conformal, jin2021sensitivity, ek2023off}, and for completeness we present the proof in full.

Following \cite{ek2023off} we introduce $\beta$ such that $0 < \beta < \smallprob$. Use~\eqref{eq:computequantile} to construct the limit 
\begin{equation}
 \loss_{\smallprob}^{\Gamma}(\setdata) = \inf \left\{ \loss : \widehat{F}(\loss;   \setloss, \overline{\weight}^{\Gamma}_\beta(\setweight)) \geq \frac{1-\smallprob}{1-\beta} \right\},
\label{eq_appx:quantile_proxy_modified}
\end{equation}
where $\overline{\weight}_\beta^{\Gamma}(\setweight)$  is defined in~\eqref{eq:weightupper_conformal}. We want to lower bound the probability of $\Loss_{\n+1} \leq \loss_{\smallprob}^{\Gamma}(\setdata)$. Note that
\begin{equation}
\begin{split}
\Prob \{\Loss_{\n+1} \leq \loss_{\smallprob}^{\Gamma}(\setdata)\} 
& = 
\Prob \{\Loss_{\n+1} \leq \loss_{\smallprob}^{\Gamma}(\setdata) \;|\; \overline{\Weight}_{\n+1}^{\Gamma} \leq \overline{\weight}_{\beta}^{\Gamma}(\setweight) \} \:
\Prob \{\overline{\Weight}_{\n+1}^{\Gamma} \leq \overline{\weight}_{\beta}^{\Gamma}(\setweight) \} \\
&\quad + 
\Prob \{\Loss_{\n+1} \leq \loss_{\smallprob}^{\Gamma}(\setdata) \; | \; \overline{\Weight}_{\n+1}^{\Gamma} > \overline{\weight}_{\beta}^{\Gamma}(\setweight) \}  \:
\Prob \{\; \overline{\Weight}_{\n+1}^{\Gamma} > \overline{\weight}_{\beta}^{\Gamma}(\setweight) \}, \\
\end{split}
\end{equation}
from the law of total probability. The second term is a lower bounded by zero, and we have
\begin{equation}
\begin{split}
\Prob \{\Loss_{\n+1} \leq \loss_{\smallprob}^{\Gamma}(\setdata)\} 
& \geq  
\Prob \{\Loss_{\n+1} \leq \loss_{\smallprob}^{\Gamma}(\setdata) \;|\; \overline{\Weight}_{\n+1}^{\Gamma} \leq \overline{\weight}_{\beta}^{\Gamma}(\setweight) \} \:
\Prob \{\overline{\Weight}_{\n+1}^{\Gamma} \leq \overline{\weight}_{\beta}^{\Gamma}(\setweight) \}.
\end{split}
\label{eq:lowerboundprob}
\end{equation}

Let us focus on the second factor in~\eqref{eq:lowerboundprob}. From the construction in~\eqref{eq:weightupper_conformal} the probability of $\overline{\Weight}_{\n+1}^{\Gamma} \leq \overline{\weight}_\beta^{\Gamma}(\setweight)$ is lower bounded by
\begin{equation}
 \Prob \{ \overline{\Weight}_{\n+1}^{\Gamma} \leq \overline{\weight}_\beta^{\Gamma}(\setweight)\} \geq 1 - \beta,
\label{eq_appx:weight_bound_guarantee}
\end{equation}
see \citet{vovk2005algorithmic}; \citet[thm.~2.1]{lei2018distribution}.

We now proceed to bound the first factor in~\eqref{eq:lowerboundprob}, i.e., $\Prob \{\Loss_{\n+1} \leq \loss_{\smallprob}^{\Gamma}(\setdata) \;|\; \overline{\Weight}_{\n+1}^{\Gamma} \leq \overline{\weight}_{\beta}^{\Gamma}(\setweight) \}$. Define the following limit
\begin{equation}
 \loss_{\smallprob}^{\Gamma}(\setloss, \overline{\Weight}_{\n+1}^{\Gamma}) = \inf \left\{ \loss : \widehat{F}(\loss;   \setloss, \overline{\Weight}_{\n+1}^{\Gamma})  \geq \frac{1-\smallprob}{1-\beta} \right\},
\label{eq_appx:quantile_proxy}
\end{equation}
where $\overline{\Weight}_{\n+1}^{\Gamma} \geq \Weight_{\n+1}$ is given in~\eqref{eq:ratio_bound}. Comparing this limit with the one defined in~\eqref{eq_appx:quantile_proxy_modified}, we see that
\begin{align}
\Prob \{\Loss_{\n+1} \leq \loss_{\smallprob}^{\Gamma}(\setdata) \;|\; \overline{\Weight}_{\n+1}^{\Gamma} \leq \overline{\weight}_{\beta}^{\Gamma}(\setweight) \}
&\geq
\Prob \{\Loss_{\n+1} \leq \loss_{\smallprob}^{\Gamma}(\setloss, \overline{\Weight}_{\n+1}^{\Gamma}) \;|\; \overline{\Weight}_{\n+1}^{\Gamma} \leq \overline{\weight}_{\beta}^{\Gamma}(\setweight) \}  \\
&= \Prob \{\Loss_{\n+1} \leq \loss_{\smallprob}^{\Gamma}(\setloss, \overline{\Weight}_{\n+1}^{\Gamma}) \} \:,
\label{eq_appx:quantile_proxy_bound}
\end{align}
whenever  $\overline{\Weight}_{\n+1}^{\Gamma} \leq \overline{\weight}_\beta^{\Gamma}(\setweight)$. The second line results from applying sample splitting, which gurantess that $\Loss_{\n+1} \leq \loss_{\smallprob}^{\Gamma}(\setloss, \overline{\Weight}_{\n+1}^{\Gamma})$ and $\overline{\Weight}_{\n+1}^{\Gamma} \leq \overline{\weight}_{\beta}^{\Gamma}(\setweight)$ are independent.

To lower bound $\Prob \{\Loss_{n+1} \leq \loss_{\smallprob}^{\Gamma}(\setloss, \overline{\Weight}_{\n+1}^{\Gamma}) \}$, we will make use of the following inequality,
\begin{equation}
\E\left[\widehat{F}(\loss_{\smallprob}^{\Gamma}; \setloss, \overline{\Weight}_{\n+1}^{\Gamma} )\right]  = \E \left[ \frac{\sum_{i \in \setloss} \underline{\Weight}_{i}^{\Gamma} \ind{\Loss_i \leq \loss_{\smallprob}^{\Gamma}}}
{\sum_{i \in \setloss}  \big[ \underline{\Weight}_{i}^{\Gamma} \ind{\Loss_i \leq \loss_{\smallprob}^{\Gamma}} + \overline{\Weight}_{i}^{\Gamma} \ind{\Loss_i > \loss_{\smallprob}^{\Gamma}} \big] + \overline{\Weight}_{\n+1}^{\Gamma}} \right]
\geq \frac{1 -\smallprob}{1 -\beta},
\label{eq_appx:expectation_upper_weight}
\end{equation}
that holds by construction.

First, define $\usetwithtest$ as an unordered set of the following elements
\begin{equation}
\big( (\X_{\nweight+1}, \Unobserved_{\nweight+1}, \Dec_{\nweight+1}, \Loss_{\nweight+1}), \dots, (\X_{m}, \Unobserved_{m}, \Dec_{m}, \Loss_{m})  ,(\X_{\n+1}, \Unobserved_{n+1}, \Dec_{\n+1}, \Loss_{\n+1}) \big).
\end{equation}
From \citet[thm.~2]{tibshirani2019conformal} we have that the out-of-sample loss $\Loss_{n+1}$ has the (conditional) cdf
\begin{equation}
\label{eq_appx:lemma_prob}
\Prob \{\Loss_{n+1} \leq \loss \; | \; \usetwithtest\}  = \sum_{i \in \usetwithtest} p_i \ind{\loss_i \leq \loss} = \frac{\sum_{i \in \usetwithtest} \weight_i \ind{\Loss_i \leq \loss}}
{\sum_{i \in \usetwithtest} \weight_i},
\end{equation}

where $\weight_i$ quantifies the distribution shift for sample $i$ using the (unobservable) ratio \eqref{eq:ratio}. Next, we build on the proof method used in \citet[thm.~2.2]{jin2021sensitivity}. Use the limit $\loss_{\smallprob}^{\Gamma}(\setloss, \overline{\Weight}_{\n+1}^{\Gamma})$ from~\eqref{eq_appx:quantile_proxy} in~\eqref{eq_appx:lemma_prob} and apply the law of total expectation to perform marginalization over $\usetwithtest$
\begin{align}
    \Prob \{\Loss_{n+1} \leq \loss_{\smallprob}^{\Gamma}(\setloss, \overline{\Weight}_{\n+1}^{\Gamma}) \} 
    & = \E \big[ \Prob \{\Loss_{n+1} \leq \loss_{\smallprob}^{\Gamma}(\setloss, \overline{\Weight}_{\n+1}^{\Gamma}) \; | \;\usetwithtest \} \big] \\
    &= \E \left[ \frac{\sum_{i \in \usetwithtest} \Weight_i \ind{\Loss_i \leq \loss_{\smallprob}^{\Gamma}(\setloss, \overline{\Weight}_{\n+1}^{\Gamma})}} {\sum_{i \in \usetwithtest} \Weight_i} \right].
    \label{eq_appx:expectation_true_weight}
\end{align}

We can now proceed to establish a lower bound for this probability. Combining~\eqref{eq_appx:expectation_upper_weight} and~\eqref{eq_appx:expectation_true_weight}, we have that
\begin{align}
    &\Prob \{\Loss_{n+1} \leq \loss_{\smallprob}^{\Gamma}(\setloss, \overline{\Weight}_{\n+1}^{\Gamma}) \} 
    -\frac{1 -\smallprob}{1 -\beta} \\
    &\geq
    \E \left[ \frac{\sum_{i \in \usetwithtest} \Weight_i \ind{\Loss_i \leq \loss_{\smallprob}^{\Gamma}}} {\sum_{i \in \usetwithtest} \Weight_i} \right] 
    - \E \left[ \frac{\sum_{i \in \setloss} \underline{\Weight}_{i}^{\Gamma} \ind{\Loss_i \leq \loss_{\smallprob}^{\Gamma}}}
{\sum_{i \in \setloss} \big[ \underline{\Weight}_{i}^{\Gamma} \ind{\Loss_i \leq \loss_{\smallprob}^{\Gamma}} + \overline{\Weight}_{i}^{\Gamma} \ind{\Loss_i > \loss_{\smallprob}^{\Gamma}} \big] + \overline{\Weight}_{\n+1}^{\Gamma}} \right] \\
    &=
    \E \left[ \frac{(*)} 
    {\left[\sum_{i \in \usetwithtest} \Weight_i\right]\left[\sum_{i \in \setloss} \big[ \underline{\Weight}_{i}^{\Gamma} \ind{\Loss_i \leq \loss_{\smallprob}^{\Gamma}} + \overline{\Weight}_{i}^{\Gamma} \ind{\Loss_i > \loss_{\smallprob}^{\Gamma}} \big] + \overline{\Weight}_{\n+1}^{\Gamma} \right] } \right],
\end{align}
where
\begin{align}
    (*) 
    =& \Bigg[\sum_{i \in \usetwithtest}\Weight_i \ind{\Loss_i \leq \loss_{\smallprob}^{\Gamma}}\Bigg]
    \Bigg[\sum_{i \in \setloss} \overline{\Weight}_{i}^{\Gamma} \ind{\Loss_i > \loss_{\smallprob}^{\Gamma}}+ \overline{\Weight}_{\n+1}^{\Gamma}\Bigg] \\
    &- \Bigg[\sum_{i \in \setloss} \underline{\Weight}_{i}^{\Gamma} \ind{\Loss_i \leq \loss_{\smallprob}^{\Gamma}}\Bigg] 
    \Bigg[\sum_{i \in \usetwithtest}\Weight_i \ind{\Loss_i > \loss_{\smallprob}^{\Gamma}} \Bigg]    \\
    \geq& \Bigg[\sum_{i \in \setloss}\Weight_i \ind{\Loss_i \leq \loss_{\smallprob}^{\Gamma}}\Bigg]
    \Bigg[\sum_{i \in \setloss} \Weight_i \ind{\Loss_i > \loss_{\smallprob}^{\Gamma}}+ W_{n+1}\Bigg] \\
    &- \Bigg[\sum_{i \in \setloss} \Weight_i \ind{\Loss_i \leq \loss_{\smallprob}^{\Gamma}}\Bigg] 
    \Bigg[\sum_{i \in \setloss}\Weight_i \ind{\Loss_i > \loss_{\smallprob}^{\Gamma}} + \Weight_{n+1} \Bigg]  \\
    =& \; 0.
\end{align}
We use the bounds provided in~\eqref{eq:ratio_bound} to derive the inequality. Hence, we arrive at a valid limit
\begin{equation}
    \Prob \{\Loss_{n+1} \leq \loss_{\smallprob}^{\Gamma}(\setloss, \overline{\Weight}_{\n+1}^{\Gamma}) \} \geq \frac{1 -\smallprob}{1 -\beta}.
\label{eq_appx:quantile_proxy_guarantee}
\end{equation}

Finally combine~\eqref{eq_appx:weight_bound_guarantee} and~\eqref{eq_appx:quantile_proxy_guarantee} to get
\begin{equation}
\begin{split}
\Prob \{\Loss_{\n+1} \leq \loss_{\smallprob}^{\Gamma}(\setdata)\} 
&\geq \Prob \{\Loss_{\n+1} \leq \loss_{\smallprob}^{\Gamma}(\setloss, \overline{\Weight}_{\n+1}^{\Gamma}) \} \:
\Prob \{\overline{\Weight}_{\n+1}^{\Gamma} \leq \overline{\weight}_{\beta}^{\Gamma}(\setweight)\} \\
&\geq 
\frac{1 -\smallprob}{1 -\beta}  \:
(1 -\beta) \\
&= 1-\smallprob.
\end{split}
\label{eq_appx:guarantee}
\end{equation}

We choose $\beta$ as in~\eqref{eq:ell_alpha} to get the tightest limit. As $\Loss_{n+1}$ is drawn from $\probpolicy(\X, \Unobserved, \Dec,\Loss|\Sampling=0)$, we can express~\eqref{eq_appx:guarantee} as shown in~\eqref{eq:certificate} for the sake of notational clarity.


\section{Numerical Experiments}
\label{sec:experiments_app}
Additional information of the numerical experiments outlined in Section \ref{sec:experiments} can be found in this supplementary material and the code used for the experiments can be accessed here \texttt{https://github.com/sofiaek/policy-evaluation-rct}. 

All experiments were conducted using Version 1.7 of the Python implementation of \xgboost\ (Apache-2.0 License). A comprehensive list of hyperparameters is available in Table \ref{tab:hyper_xgboost} for the synthetic case and Table~\ref{tab:hyper_xgboost_nhanes} for the NHANES case. The hyperparameters were selected through a random search involving 200 runs, employing 5-fold cross-validation with the F1 score as the optimization metric. All experiments were performed on a laptop with the following specifications: Intel Core i7-8650 CPU @ 1.9GHz, 16 GB DDR4 RAM, and Windows 10 Pro 64-bit operating system. The experiments utilized only the CPU. The total time required to run all the experiments was approximately half an hour.

\begin{table}
    \caption{Hyperparameters used for \xgboost{} in Section~\ref{sec:synthetic-data}.} 
    \label{tab:hyper_xgboost}
    \medskip
    \centering
    \begin{tabular}{lc}
        \toprule
        Parameter & Value \\
        \midrule
        \texttt{n\_estimators} & $100$   \\ 
        \texttt{max\_depth} & $2$   \\  
        \texttt{learning\_rate} & $0.05$   \\
        \texttt{objective} & \texttt{binary:logistic}    \\ 
        \texttt{min\_child\_weight} & $1$   \\  
        \texttt{subsample} & $0.6$   \\  
        \texttt{colsample\_bytree} & $0.8$   \\  
        \texttt{colsample\_bylevel} & $0.4$   \\  
        \texttt{scale\_pos\_weight} & $n_{s=0}/n_{s=1}$   \\         
        \bottomrule
    \end{tabular}
\end{table}

\begin{table}
    \caption{Hyperparameters used for \xgboost{} in Section~\ref{sec:semi-real data}.} 
    \medskip
    \centering
    \begin{tabular}{lc}
        \toprule
        Parameter & Value \\
        \midrule
        \texttt{n\_estimators} & $50$   \\ 
        \texttt{max\_depth} & $1$   \\  
        \texttt{learning\_rate} & $0.2$   \\
        \texttt{objective} & \texttt{binary:logistic}    \\ 
        \texttt{min\_child\_weight} & $1$   \\  
        \texttt{subsample} & $0.4$   \\  
        \texttt{colsample\_bytree} & $0.7$   \\  
        \texttt{colsample\_bylevel} & $0.8$   \\  
        \texttt{scale\_pos\_weight} & $n_{s=0}/n_{s=1}$   \\         
        \bottomrule
    \end{tabular}
    \label{tab:hyper_xgboost_nhanes}
\end{table}

\subsection{Synthetic data}
We extend the experiments in Section~\ref{sec:synthetic-data} to include two extra target populations \texttt{C}, respective \texttt{D}. The full list of parameters used in~\eqref{eq:sim_covariates} are given in Table~\ref{tab:simulations_appendix}.

\begin{table}
    \caption{Means and variances of covariate distribution $\prob(\X, \Unobserved|\Sampling)$ in~\eqref{eq:sim_covariates}.} 
    \medskip
    \centering
    \begin{tabular}{ccccccc}
        \toprule
        Population & $\mu_{0, \Sampling}$ & $\mu_{1, \Sampling}$ & $\mu_{\Unobserved, \Sampling}$ & $\sigma_{0, \Sampling}^2$ & $\sigma_{1, \Sampling}^2$ & $\sigma_{\Unobserved, \Sampling}^2$ \\
        \midrule
        \texttt{A} $(\Sampling=0)$ & 0.5 & 0.5 & 0.5 & 1.0 & 1.0 & 1.0   \\ 
        \texttt{B} $(\Sampling=0)$ & 0.0 & 0.5 & 0.0 & 1.25& 1.5 & 1.25 \\
        \texttt{C} $(\Sampling=0)$ & 0.0 & 0.0 & 0.0 & 1.5& 1.5 & 1.5   \\ 
        \texttt{D} $(\Sampling=0)$ & 0.25 & 0.25 & 0.25 & 1.0 & 0.25 & 0.5   \\ 
        \midrule
        \texttt{Trial} $(\Sampling=1)$ & 0.0 & 0.0 & 0.0 & 1.0 & 1.0 & 1.0   \\
        \bottomrule
    \end{tabular}
    \label{tab:simulations_appendix}
\end{table}

\begin{figure}
\centering
    \begin{subfigure}[b]{0.9\textwidth}
        \centering
        \includegraphics[width=1.0\linewidth]{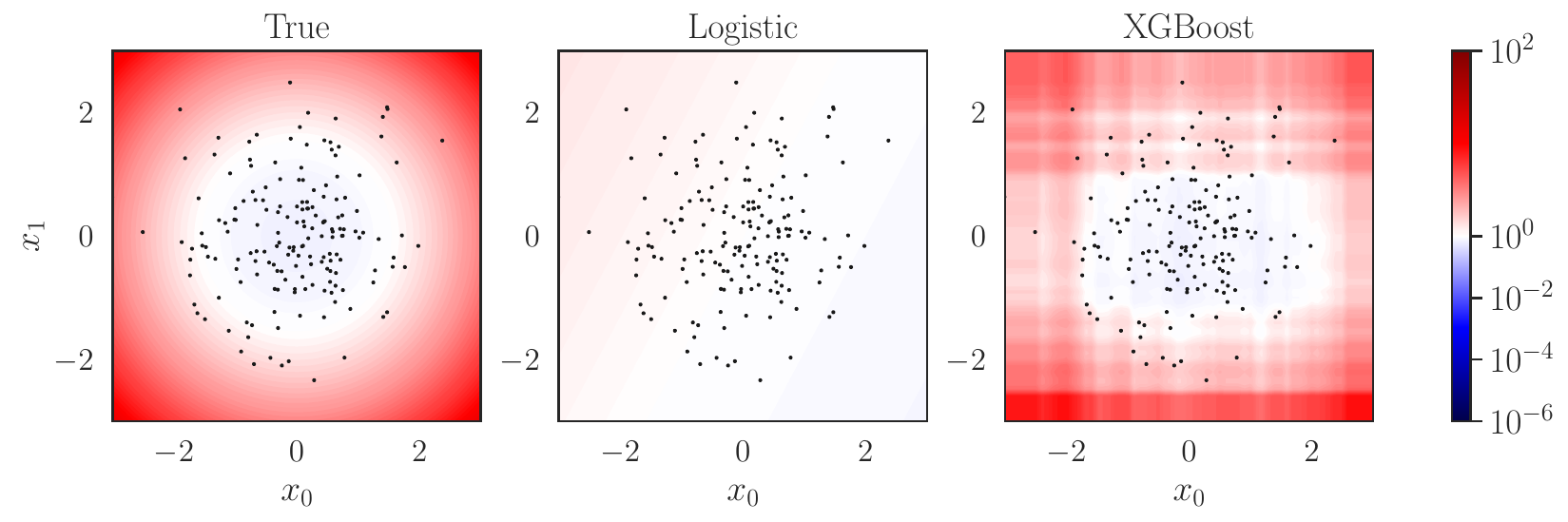}
        \caption{Sampling odds for for target population \texttt{C}.}
        \label{fig:synth_odds_testC}
    \end{subfigure}   
    \vskip\baselineskip
    \begin{subfigure}[b]{0.9\textwidth}
        \centering
         \includegraphics[width=1.0\linewidth]{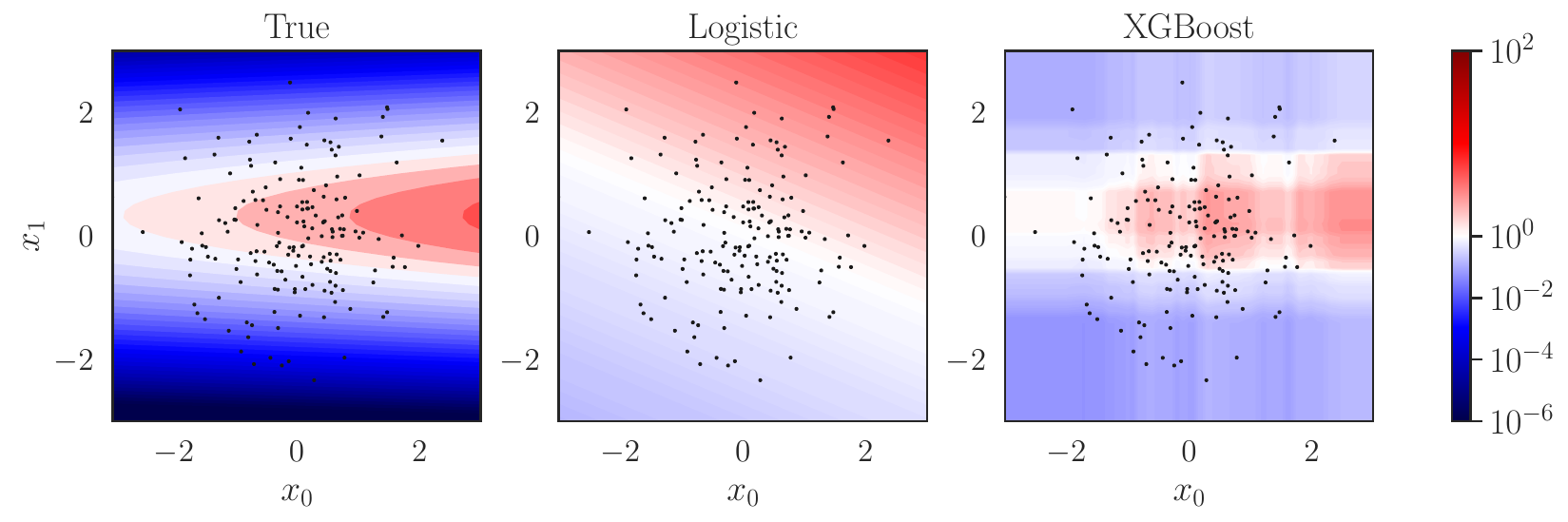}
        \caption{Sampling odds for target population \texttt{D}.}
         \label{fig:synth_odds_testD}
    \end{subfigure}
    \caption{Odds $\prob(\Sampling=0|\X)/\prob(\Sampling=1|\X)$ compared with nominal odds obtained from logistic and \xgboost{} models $\probhat(\Sampling|\X)$. The dots are a random subsample of the trial samples.}
    \label{fig:synth_odds_appx}
\end{figure}

To illustrate the generality of the proposed methodology, we consider two different fitted models $\probhat(\Sampling|\X)$: a logistic model, which is conventionally used in the causal inference literature, and the more flexible tree-based ensemble model trained by \xgboost{}. Figure~\ref{fig:synth_odds_appx} compares nominal sampling odds obtained from the fitted models with the unknown odds, $\prob(\Sampling=0|\X)/\prob(\Sampling=1|\X)$, for the target populations \texttt{C}, respective \texttt{D}. This corresponds to a case without unmeasured selection factors $\Unobserved$. In all these cases, the logistic model is misspecified and miscalibrated while the \xgboost\ model provides odds that resemble the true ones. Note that the proposed algorithm can handle any model if~\eqref{eq:Gamma_divergence} is satisfied. However, a well-performing model is generally required to achieve a small $\Gamma$ and for a meaningful benchmark of the upper value of $\Gamma$.  

\begin{figure}
    \begin{subfigure}[b]{0.49\textwidth}
        \centering
        \includegraphics[width=0.975\linewidth]{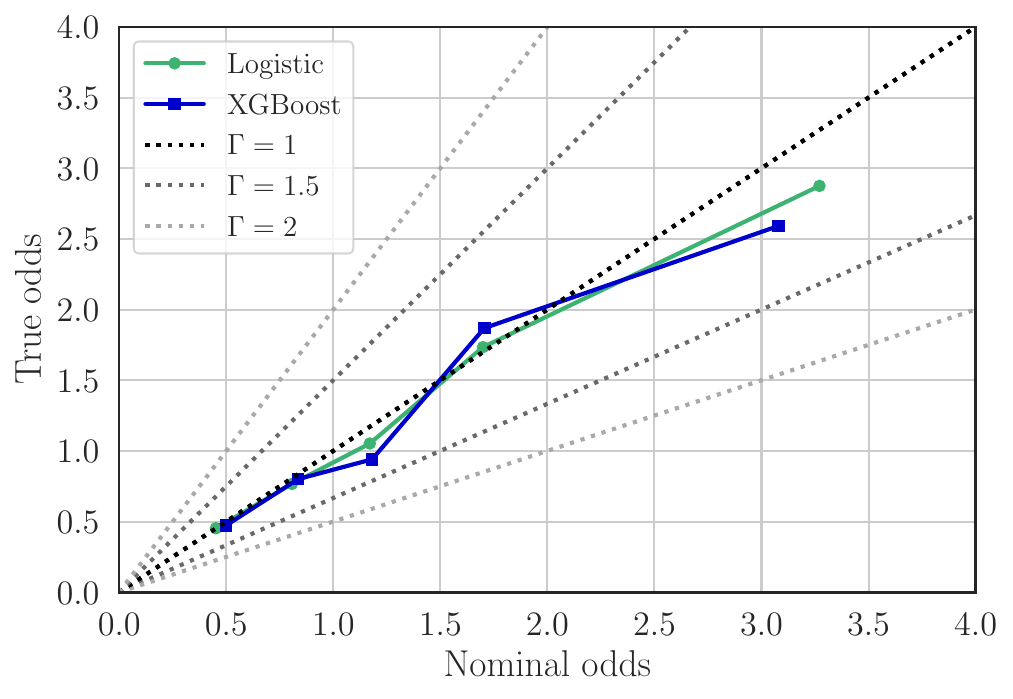}
        \caption{Reliability diagram for target population \texttt{A}.}
        \label{fig:reliability_testA}
    \end{subfigure}
    \begin{subfigure}[b]{0.49\textwidth}
        \centering
        \includegraphics[width=0.975\linewidth]{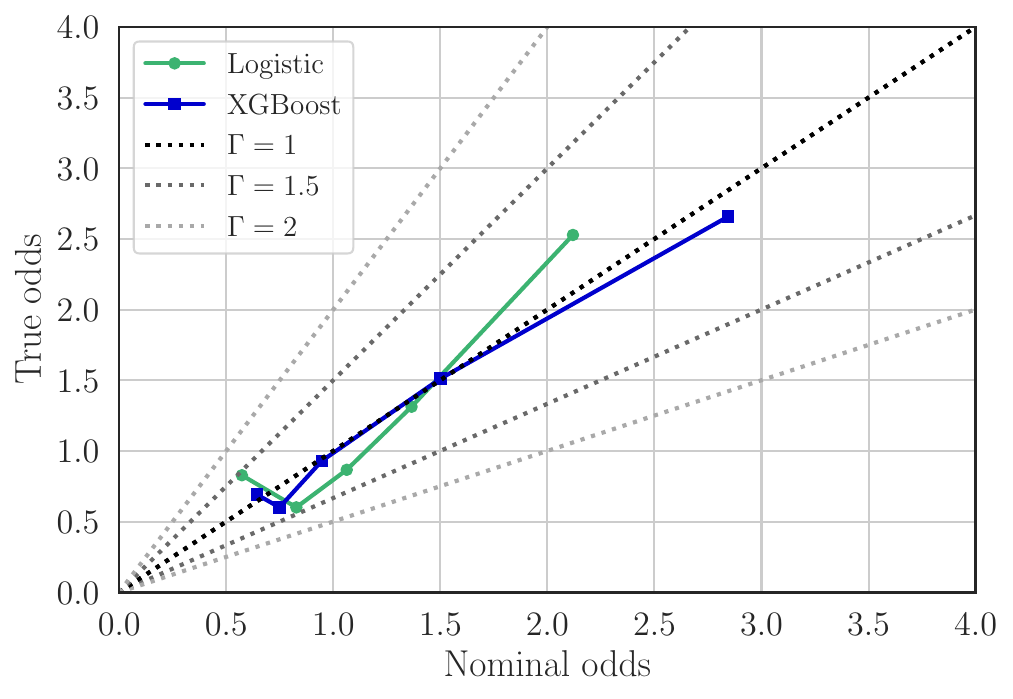}
        \caption{Reliability diagram for target population \texttt{B}.}
        \label{fig:reliability_testB_appx}
    \end{subfigure}
    \vskip\baselineskip
    \begin{subfigure}[b]{0.49\textwidth}
        \centering
        \includegraphics[width=0.975\linewidth]{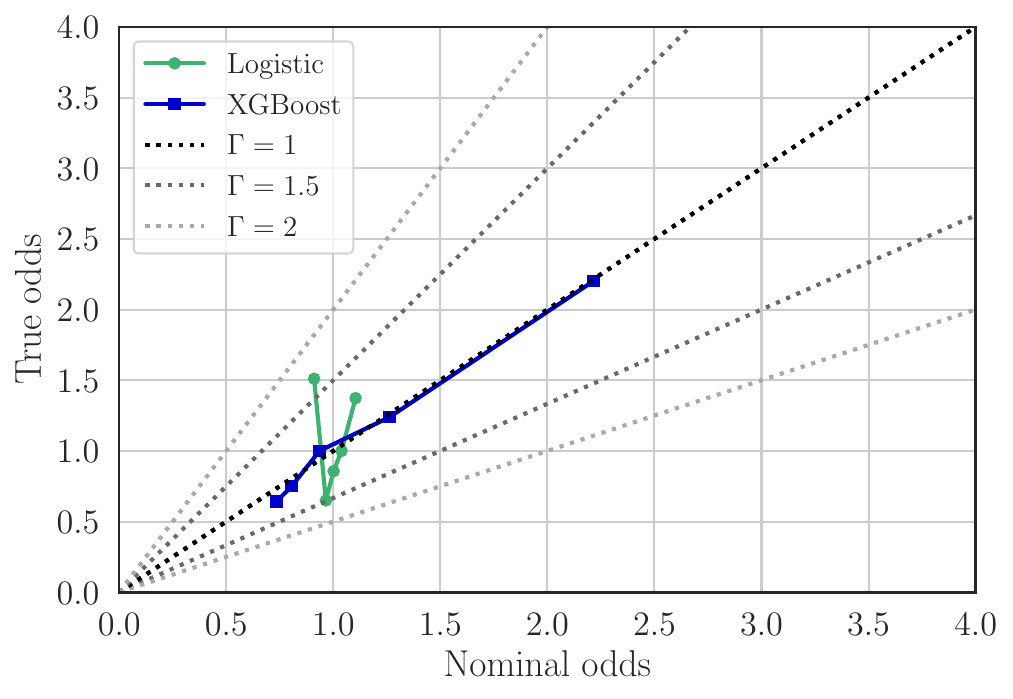}
        \caption{Reliability diagram for target population \texttt{C}.}
        \label{fig:reliability_testC}
    \end{subfigure}
        \begin{subfigure}[b]{0.49\textwidth}
        \centering
        \includegraphics[width=0.975\linewidth]{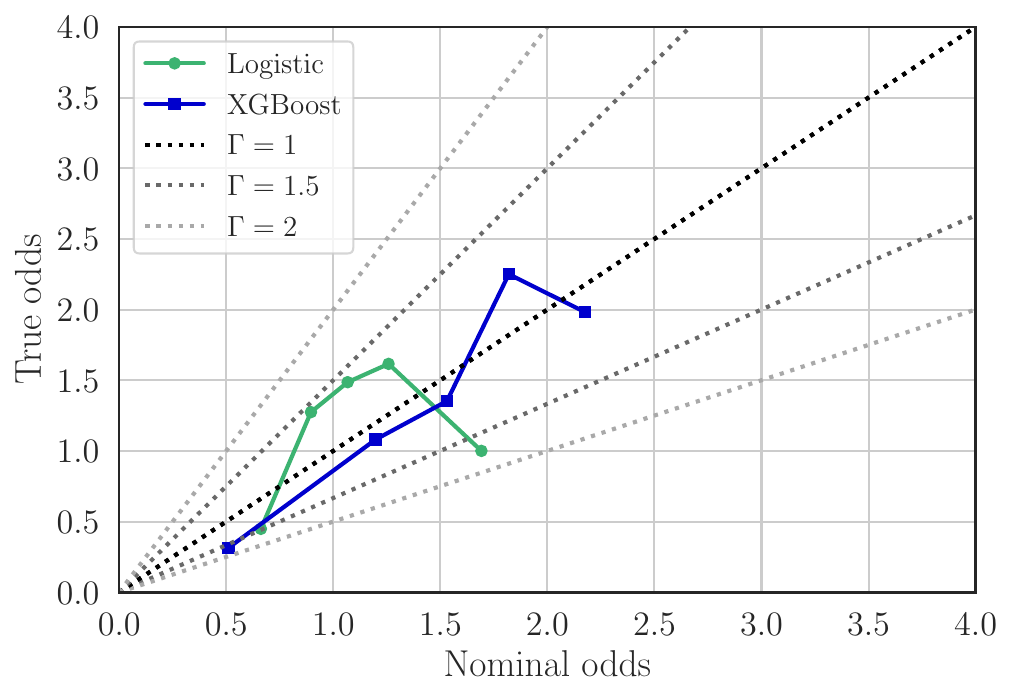}
        \caption{Reliability diagram for target population \texttt{D}.}
        \label{fig:reliability_testD}
    \end{subfigure}
    \caption{Reliability diagram of the observed odds against the average predicted nominal odds obtained from logistic and \xgboost{}-trained models $\probhat(\Sampling|\X)$.}
    \label{fig:reliability_appx}
\end{figure}

In Figure~\ref{fig:reliability_appx}, we use the reliability diagram technique described in subsection~\ref{sec:benchmarking_gamma} to assess and compare the performance of logistic and \xgboost{} models of the nominal odds, $\widehat{\odds}(\X)$, for the synthetic data. For all numerical experiments, we use 5 bins. For target population \texttt{A} in Figure~\ref{fig:reliability_testA}, both models are close to the diagonal and it is reasonable to believe that they are flexible enough to model the odds. For the other three populations, \texttt{B}, \texttt{C} and \texttt{D}, in Figure~\ref{fig:reliability_testB_appx}, \ref{fig:reliability_testC} respective \ref{fig:reliability_testD} it is evident that the \xgboost{} model shows a closer alignment with the diagonal compared to the logistic model. This is in line with the results in Figure~\ref{fig:synth_odds_test2} and~\ref{fig:synth_odds_appx} where the \xgboost{} model is visually closer to the true model.

\begin{figure}
\centering
    \begin{subfigure}[b]{0.49\textwidth}
        \centering
        \includegraphics[width=1.0\linewidth]{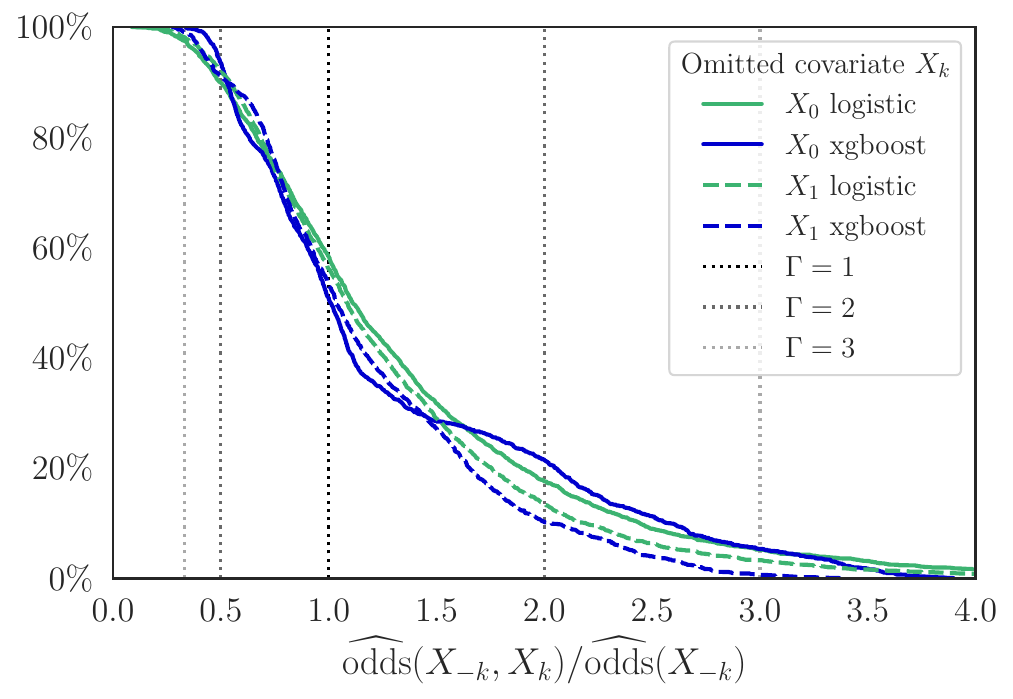}
        \caption{Benchmarking for population \texttt{A}.}
        \label{fig:synth_calibration_appx_testA}
    \end{subfigure}   
    \begin{subfigure}[b]{0.49\textwidth}
        \centering
        \includegraphics[width=1.0\linewidth]{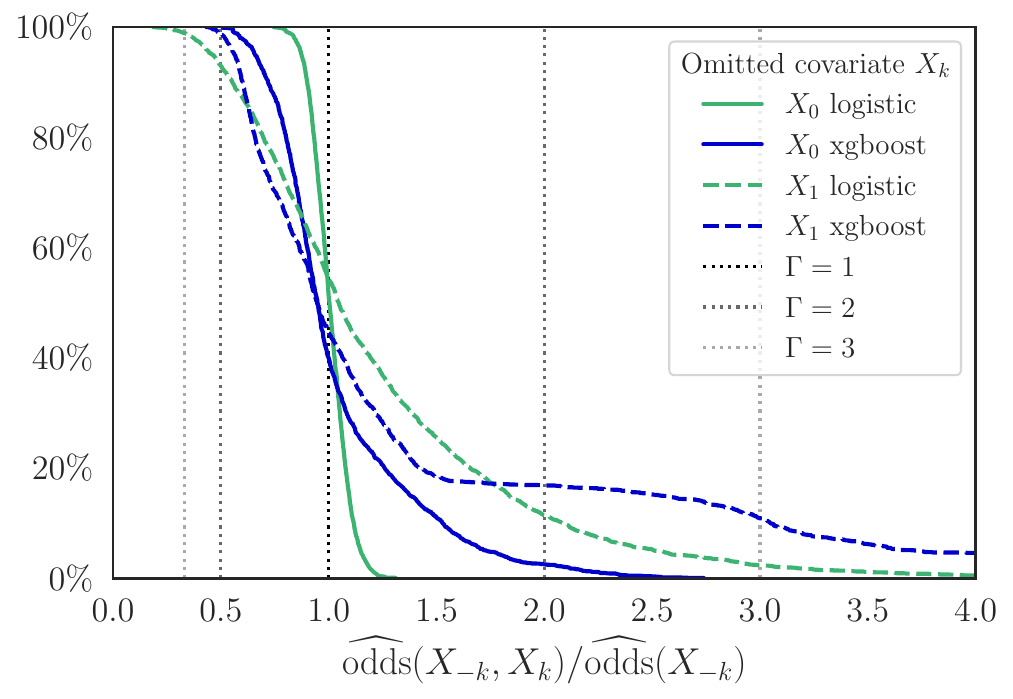}
        \caption{Benchmarking for population \texttt{B}.}
        \label{fig:synth_calibration_appx_testB}
    \end{subfigure} 
     \vskip\baselineskip
    \begin{subfigure}[b]{0.49\textwidth}
        \centering
        \includegraphics[width=1.0\linewidth]{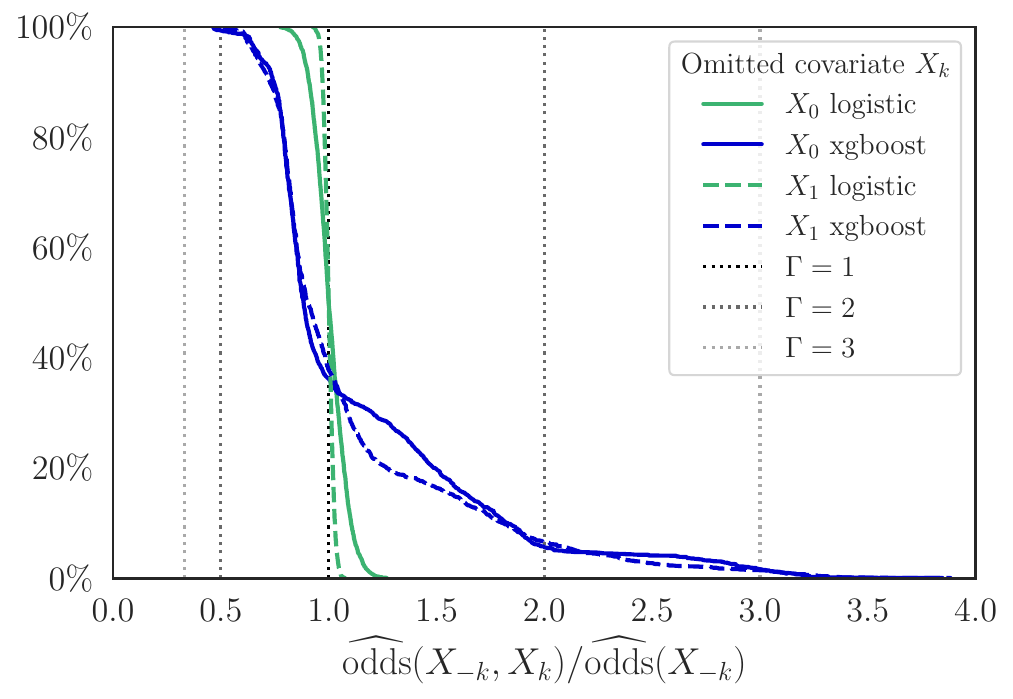}
        \caption{Benchmarking for population \texttt{C}.}
        \label{fig:synth_calibration_appx_testC}
    \end{subfigure}   
    \begin{subfigure}[b]{0.49\textwidth}
        \centering
        \includegraphics[width=1.0\linewidth]{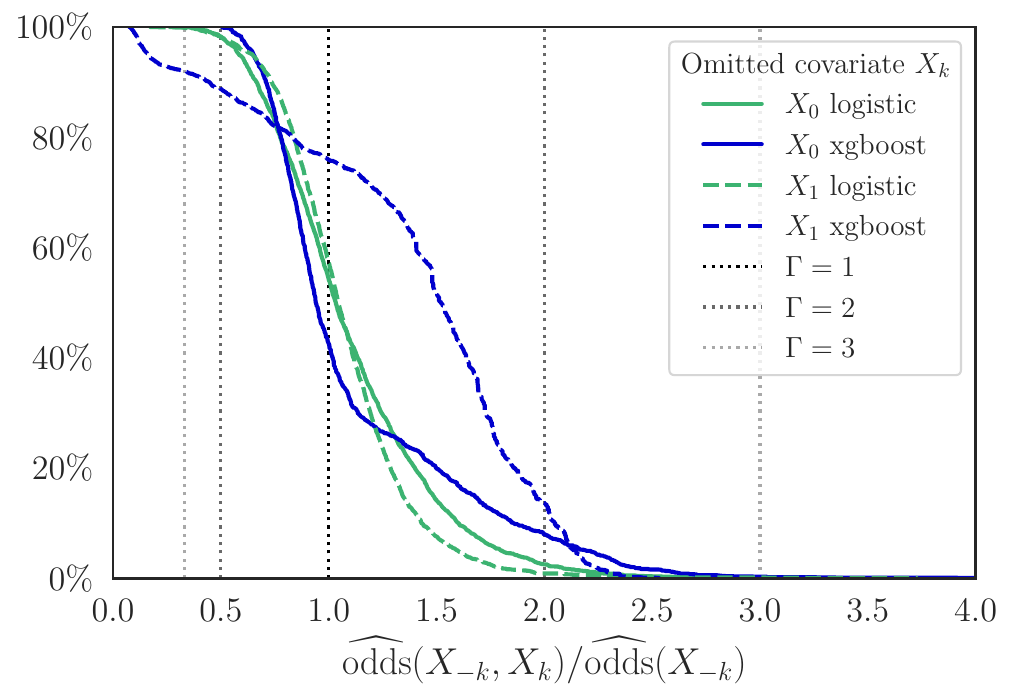}
        \caption{Benchmarking for population \texttt{D}.}
        \label{fig:synth_calibration_appx_testD}
    \end{subfigure}   
    \caption{Benchmarking the degree of miscalibration $\Gamma$ using omitted covariates.}
    \label{fig:synth_calibration_appx}
\end{figure}

We now use the observed covariates to benchmark appropriate upper bounds for $\Gamma$ for all target populations as described in subsection~\ref{sec:benchmarking_gamma}. Figure~\ref{fig:synth_calibration_appx_testA} shows the evaluation with respect to population \texttt{A}. Assuming that the unmeasured selection factors $\Unobserved$ are of similar strength as the weakest covariate, $\X_1$, a $\Gamma$ value of $2.5$ could be a credible choice, as it covers $90\%$ of the odds ratios. Figure~\ref{fig:synth_calibration_appx_testB} shows the same evaluation with respect to population \texttt{B}. A $\Gamma$ value of $1.2$ could be a credible choice for the logistic regression model and a $\Gamma$ value of $2$ could be appropriate for the \xgboost\ model. However, from Figure~\ref{fig:synth_odds_test2} we know that the logistic model is misspecified in this scenario. For population \texttt{C} in Figure~\ref{fig:synth_calibration_appx_testC}, $1$ and $2.5$ seem to be suitable $\Gamma$ values for the logistic regression model respective the \xgboost\ model. Similarly, for population \texttt{D} in Figure~\ref{fig:synth_calibration_appx_testD}, $1.5$ and $2.2$ could be suitable values for the two models. The logistic model is again misspecified in populations \texttt{C} and \texttt{D}.

\begin{figure}
    \centering
    \begin{subfigure}[b]{0.975\textwidth}
        \centering
        \includegraphics[width=0.48\linewidth]{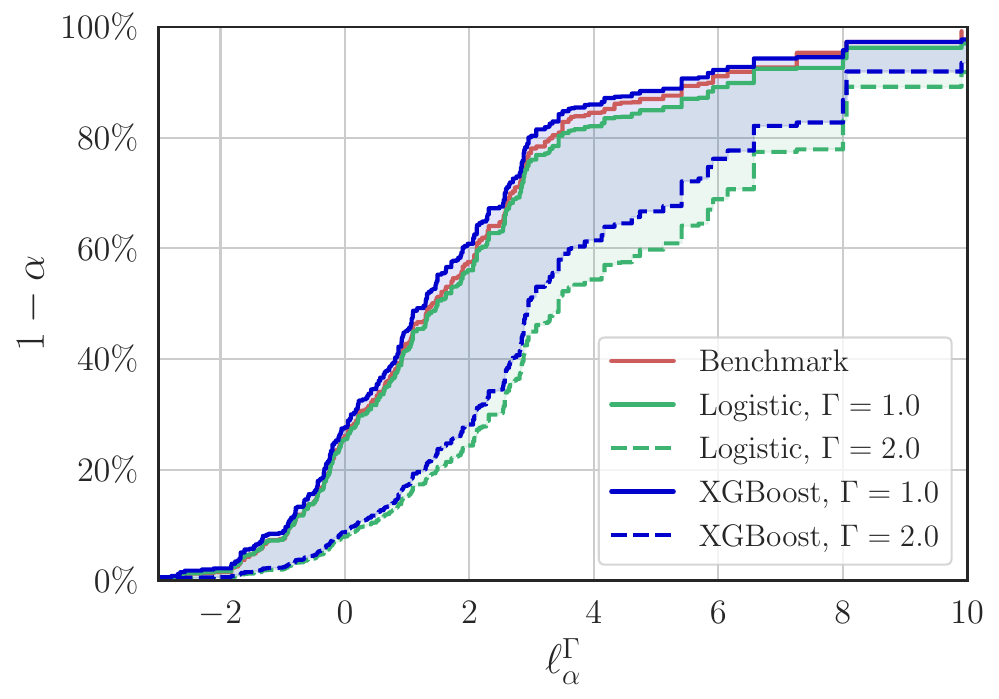} 
        \hfill
        \includegraphics[width=0.48\textwidth]{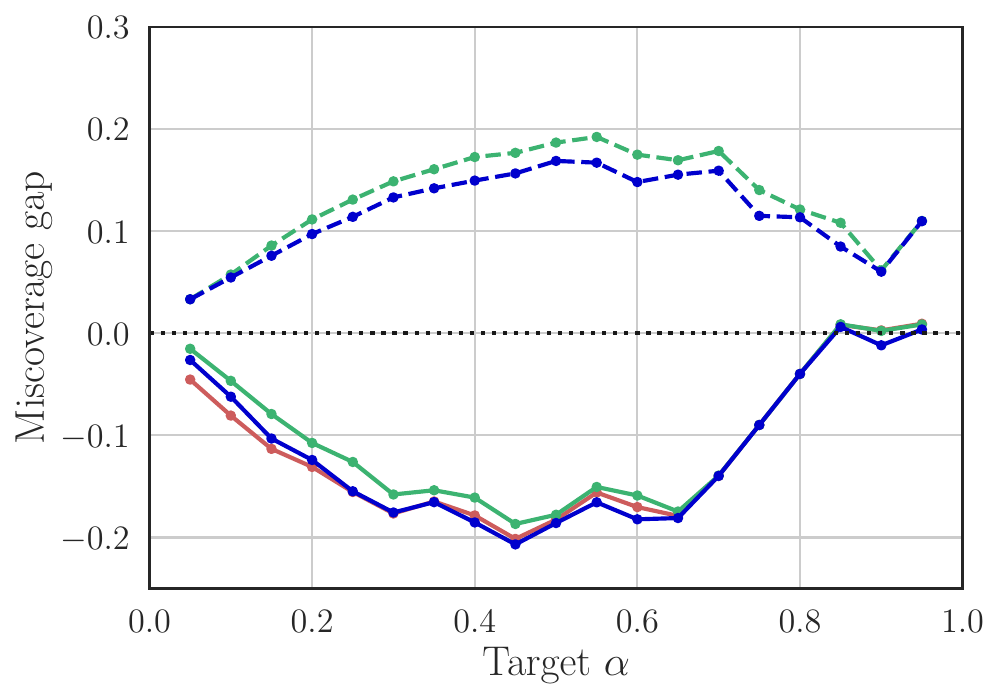}
        \caption{Limit curve and miscoverage gap for target population \texttt{A}.}
        \label{fig:synth_limit_testA}
    \end{subfigure}
    \vskip\baselineskip
    \begin{subfigure}[b]{0.975\textwidth}
        \centering
        \includegraphics[width=0.48\linewidth]{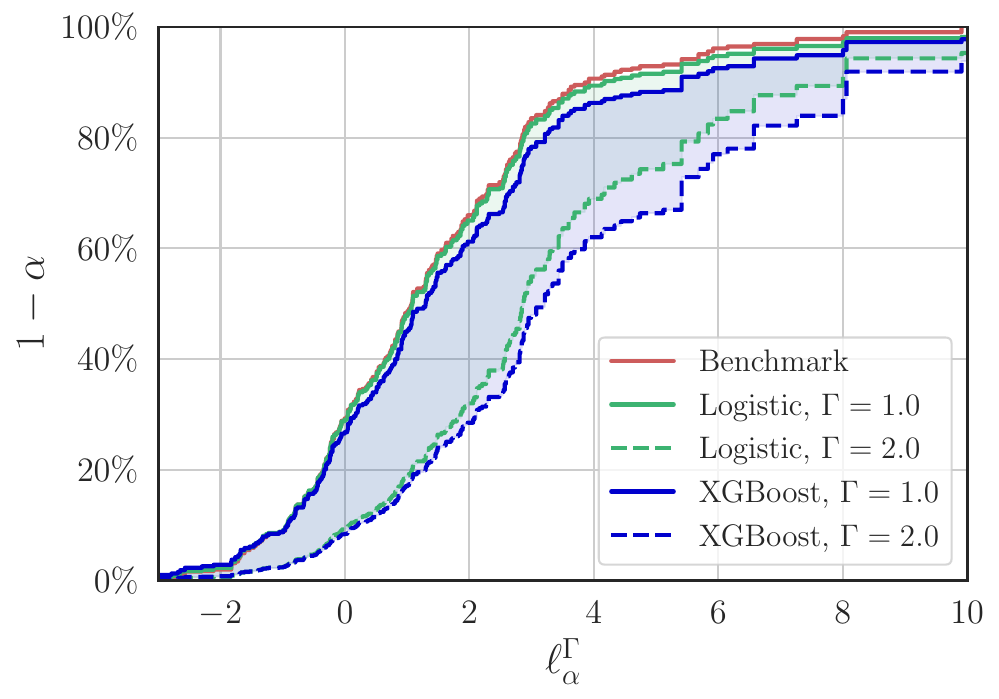} 
        \hfill
        \includegraphics[width=0.48\textwidth]{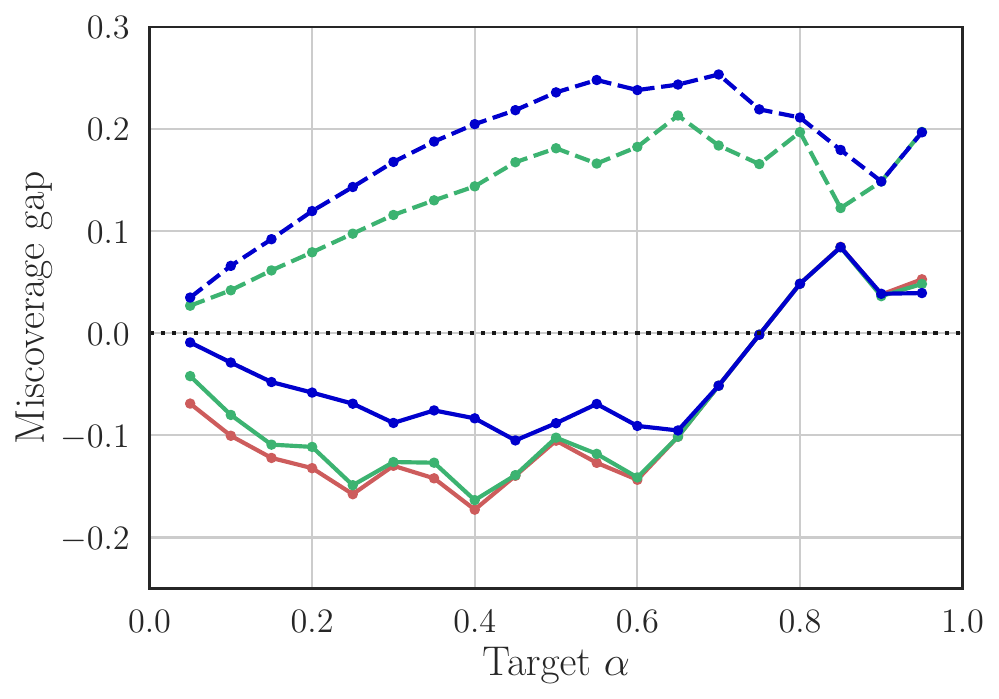}
        \caption{Limit curve and miscoverage gap for target population \texttt{B}.}
        \label{fig:synth_limit_testB}
    \end{subfigure}
    \vskip\baselineskip
    \begin{subfigure}[b]{0.975\textwidth}
        \centering
        \includegraphics[width=0.48\linewidth]{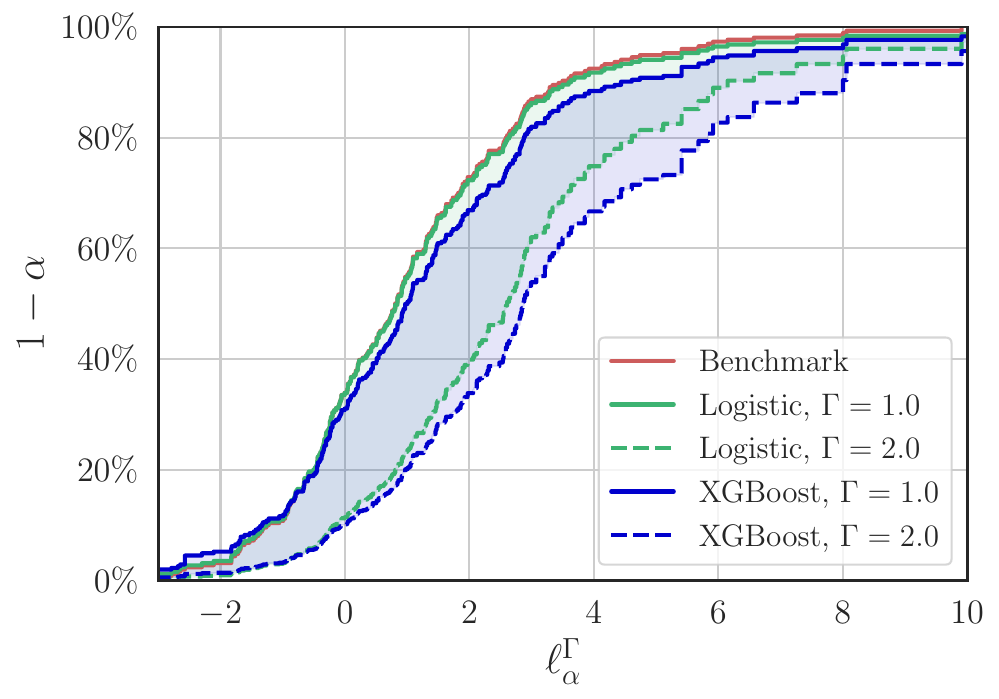} 
        \hfill
        \includegraphics[width=0.48\textwidth]{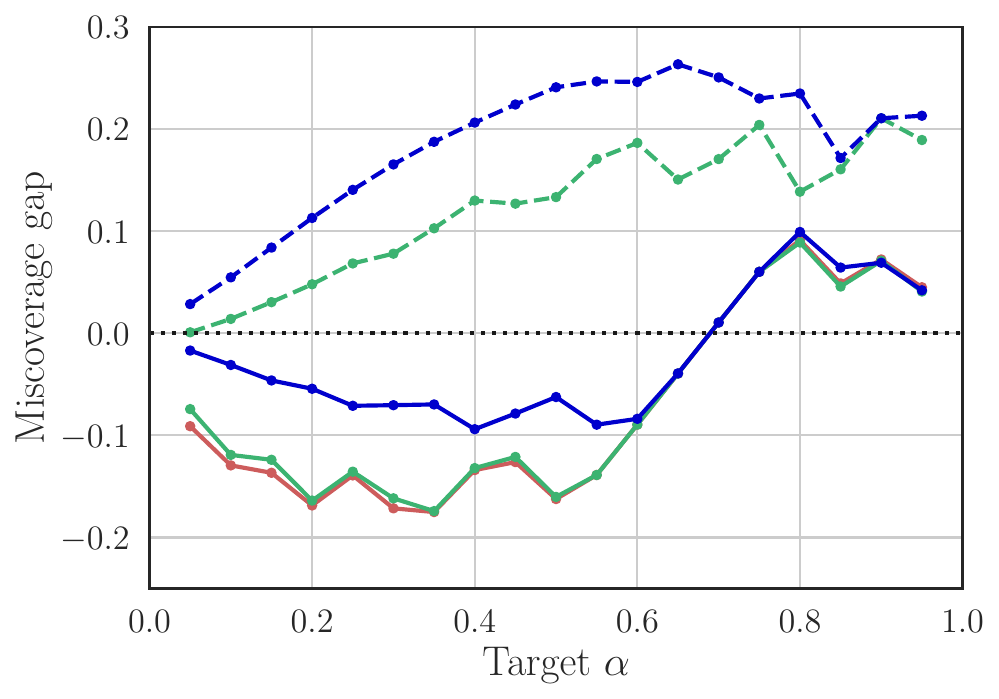}
        \caption{Limit curve and miscoverage gap for target population \texttt{C}.}
        \label{fig:synth_limit_testC}
    \end{subfigure}
    \vskip\baselineskip
    \begin{subfigure}[b]{0.975\textwidth}
        \centering
        \includegraphics[width=0.48\linewidth]{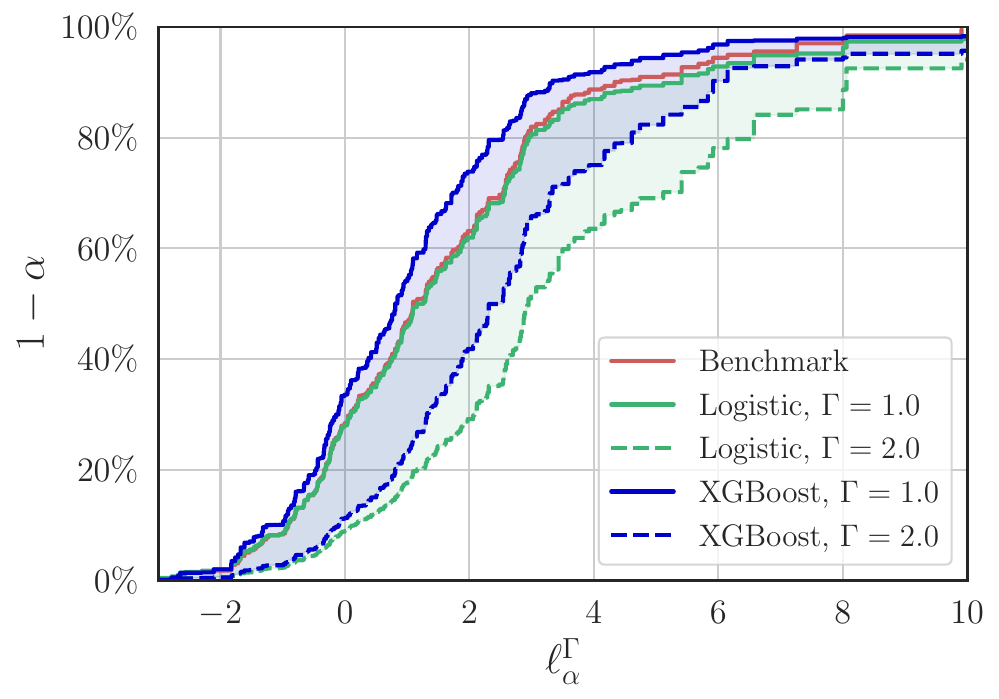} 
        \hfill
        \includegraphics[width=0.48\textwidth]{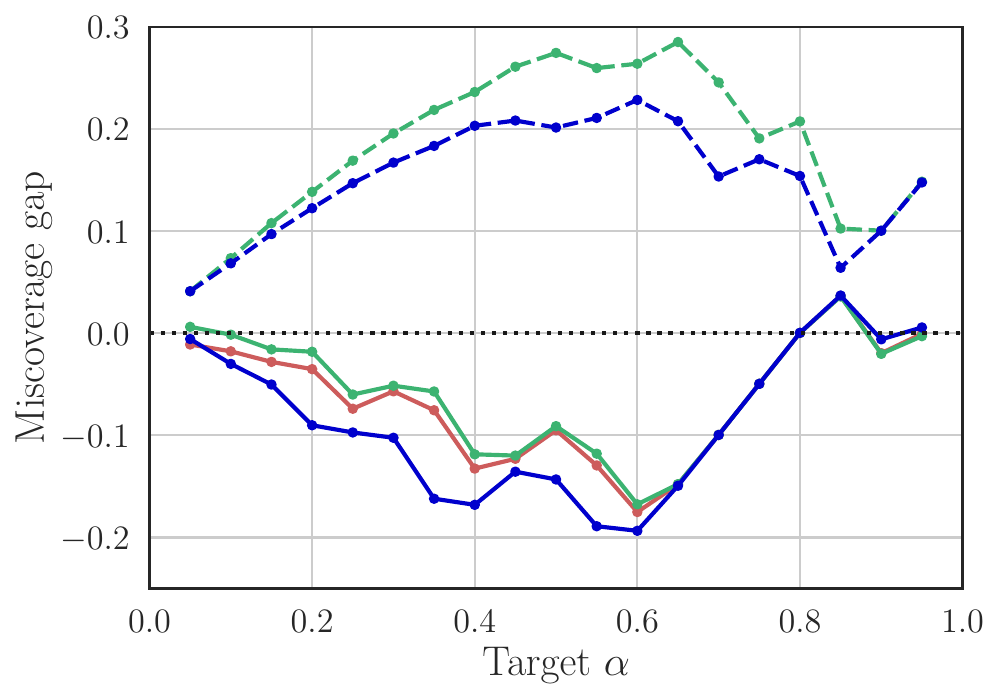}
        \caption{Limit curve and miscoverage gap for target population \texttt{D}.}
       \label{fig:synth_limit_testD}
    \end{subfigure}
    \caption{Evaluating `treat all'  policy $\policy_1$ for different target populations with degrees of miscalibration $\Gamma \in  [1, 2]$.}
    \label{fig:synth_limit_appx}
\end{figure}

In Figure~\ref{fig:synth_limit_appx}, we use the limit curves for the benchmark in~\eqref{eq:quantile_empipw} and the proposed method to evaluate the out-of-sample loss of the `treat all' policy $\policy_1$, i.e. 
\begin{equation}
    \prob_{\pi_1}(\Dec = 0 \; | \; \X) = 1.
\end{equation} 
Figure~\ref{fig:synth_limit_testA} shows the evaluation with respect to population \texttt{A}. When $\Gamma =1$ all the limit curves are similar. The curves for the proposed method also illustrate increasing the credibility of the models results in less informative inferences. However, their informativeness stays above 90\% for odds miscalibration degrees $\Gamma \in [1,2]$. We also evaluate the miscoverage gap of the curves and observe that the benchmark and the limit curves for $\Gamma =1$ have a negative miscoverage gap. As the degree of miscalibration $\Gamma$ increases to $2$, the limit curves exhibit positive miscoverage gaps, where \xgboost{} results in slightly less conservative inferences than the logistic model does. We continue with population \texttt{B} and \texttt{C} in Figure~\ref{fig:synth_limit_testB} respective \ref{fig:synth_limit_testC}. The limit curves derived for the baselines closely align with the curves modelled using logistic regression when $\Gamma =1$, but is consistently lower than the curves modelled using \xgboost{}. For the certified curves the informativeness stays above 90\% for odds miscalibration degrees $\Gamma \in [1,2]$. For the miscoverage gap, the baselines and the limit curves for $\Gamma=1$ are invalid. As the degree of miscalibration $\Gamma$ increases to $2$, all limit curves indicate positive miscoverage gaps. In Figure~\ref{fig:synth_limit_testD}, we evaluate the same for population \texttt{D}. The limit curves modelled using the baseline and logistic regression when $\Gamma = 1$ infer consistently higher losses than the curve modelled using \xgboost{}. For the curve using the logistic model the informativeness stays above 90\% for odds miscalibration degrees $\Gamma \in [1,2]$. The same figure for the \xgboost{} model is approximately 95\%. For the miscoverage gap, the baseline and the limit curves for $\Gamma=1$ are invalid. Again, as the degree of miscalibration $\Gamma$ increases to $2$, all limit curves indicate positive miscoverage gaps.

\begin{figure}
    \begin{subfigure}[b]{0.49\textwidth}
        \centering
        \includegraphics[width=0.975\linewidth]{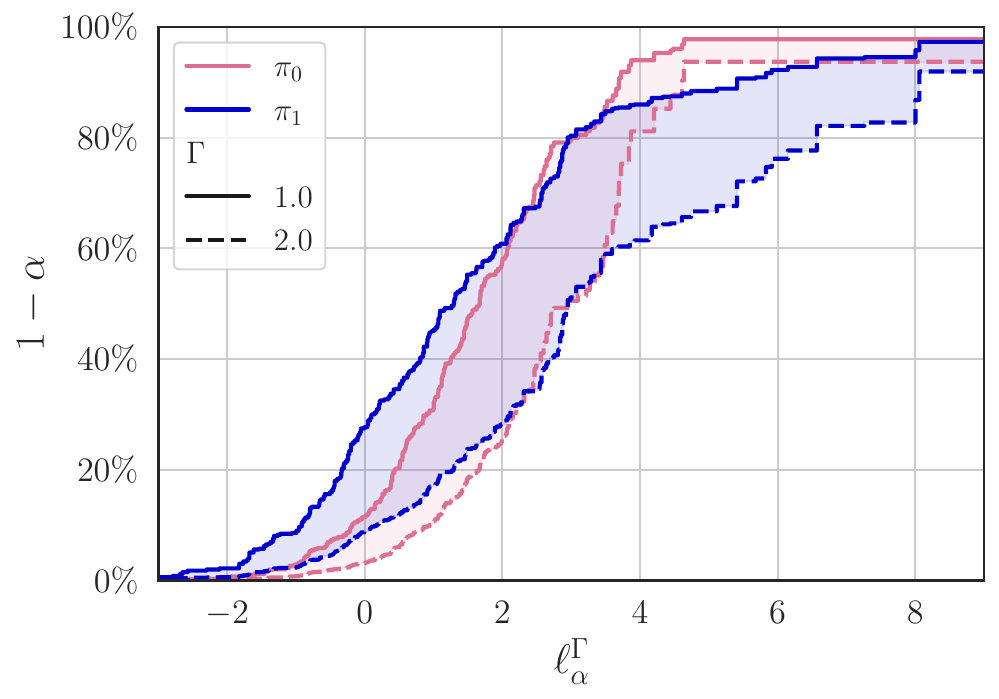}
        \caption{Limit curves for target population \texttt{A}.}
        \label{fig:compare_policyA:appx}
    \end{subfigure}
    \begin{subfigure}[b]{0.49\textwidth}
        \centering
        \includegraphics[width=0.975\linewidth]{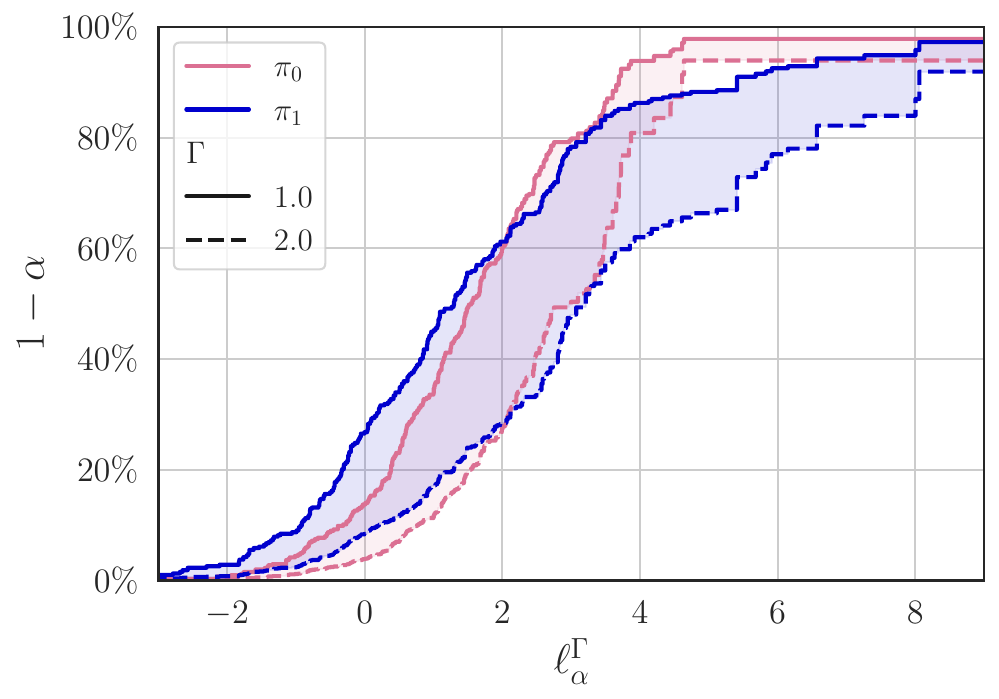}
        \caption{Limit curves for target population \texttt{B}.}
        \label{fig:compare_policyB:appx}
    \end{subfigure}
    \vskip\baselineskip
    \begin{subfigure}[b]{0.49\textwidth}
        \centering
        \includegraphics[width=0.975\linewidth]{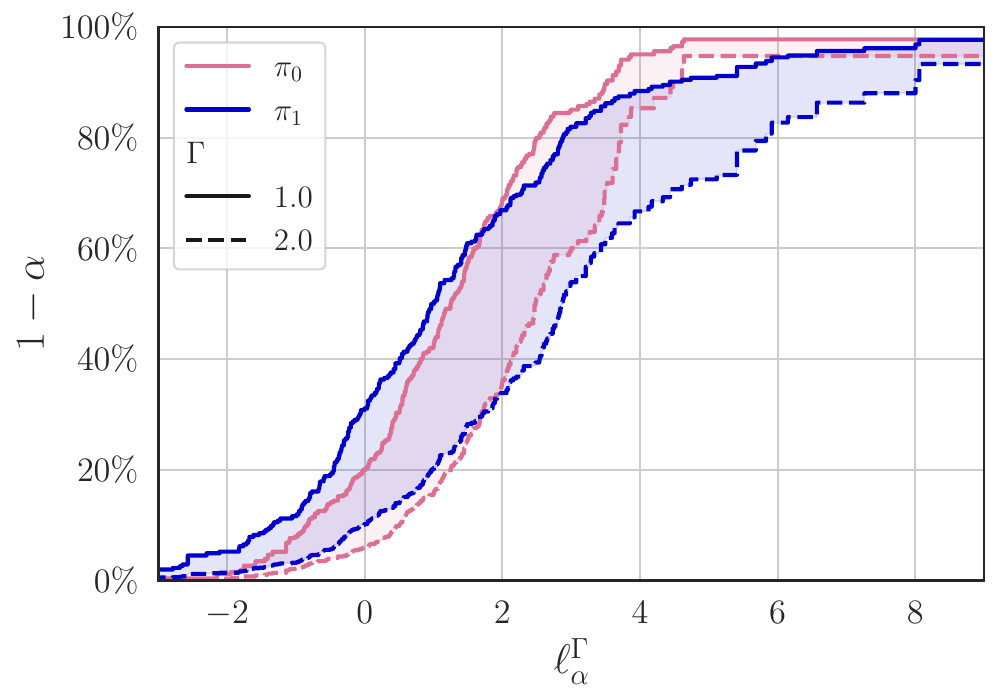}
        \caption{Limit curves for target population \texttt{C}.}
        \label{fig:compare_policyC:appx}
    \end{subfigure}
        \begin{subfigure}[b]{0.49\textwidth}
        \centering
        \includegraphics[width=0.975\linewidth]{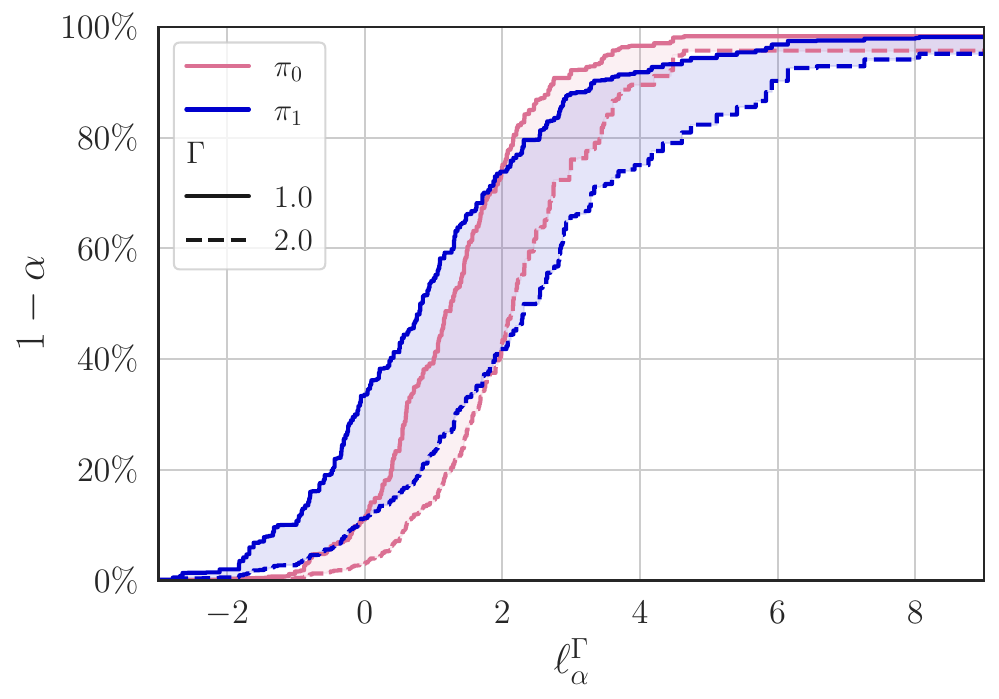}
        \caption{Limit curves for target population \texttt{D}.}
        \label{fig:compare_policyD:appx}
    \end{subfigure}
    \caption{Limit curves for $\pi_0$ and $\pi_1$ for different target populations certified for $\Gamma \in [1,2]$.}
    \label{fig:compare_policy:appx}
\end{figure}

We now turn to comparing the `treat all' policy with a `treat none' policy, i.e., 
\begin{equation}
    \prob_{\pi_0}(\Dec = 0 \; | \; \X) = 1,  
\end{equation}
for population \texttt{A}. Their expected losses are estimated using~\eqref{eq:ipsw} as $ V^{\policy_0}_{\text{IPSW}} = 1.64$  and  $V^{\policy_1}_{\text{IPSW}} = 1.54$, respectively. This evaluation suggests that $\policy_1$ is preferable to $\policy_0$. However, the limit curves presented in Figure~\ref{fig:compare_policyA:appx} provide a more detailed picture in terms of out-of-sample losses: the tail losses certified for the  `treat none' policy $\policy_0$ are lower than those certified for $\policy_1$. This illustrates the cautionary principle built into the policy evaluations. Similar results are observed for target population \texttt{D}. While for target populations \texttt{B} and \texttt{C}, both $ V^{\policy_0}_{\text{IPSW}}$ and  $V^{\policy_1}_{\text{IPSW}}$ exhibit comparable sizes, the proposed limit curves still offer a more detailed understanding of out-of-sample losses.  

\begin{figure}
    \begin{subfigure}[b]{0.49\textwidth}
        \centering
        \includegraphics[width=0.975\linewidth]{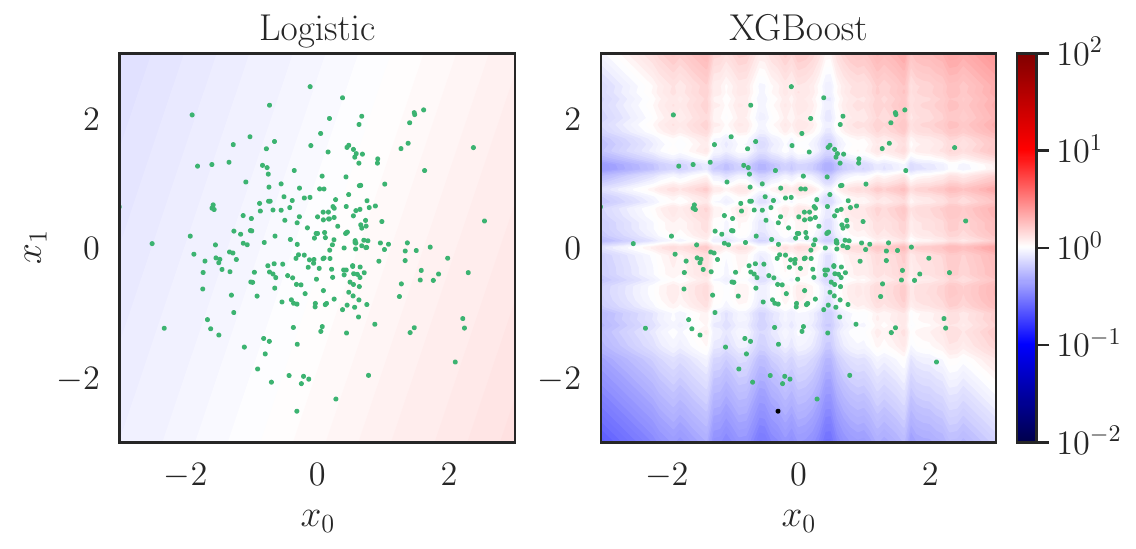}
        \caption{Degree of miscalibration for target population \texttt{A}.}
    \end{subfigure}
    \begin{subfigure}[b]{0.49\textwidth}
        \centering
        \includegraphics[width=0.975\linewidth]{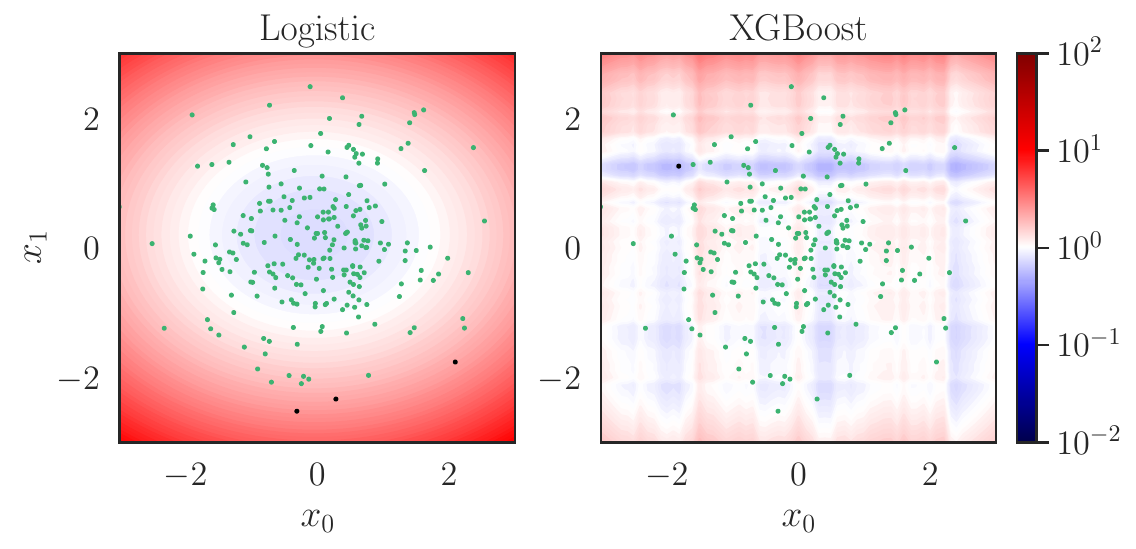}
        \caption{Degree of miscalibration for target population \texttt{B}.}
    \end{subfigure}
    \vskip\baselineskip
    \begin{subfigure}[b]{0.49\textwidth}
        \centering
        \includegraphics[width=0.975\linewidth]{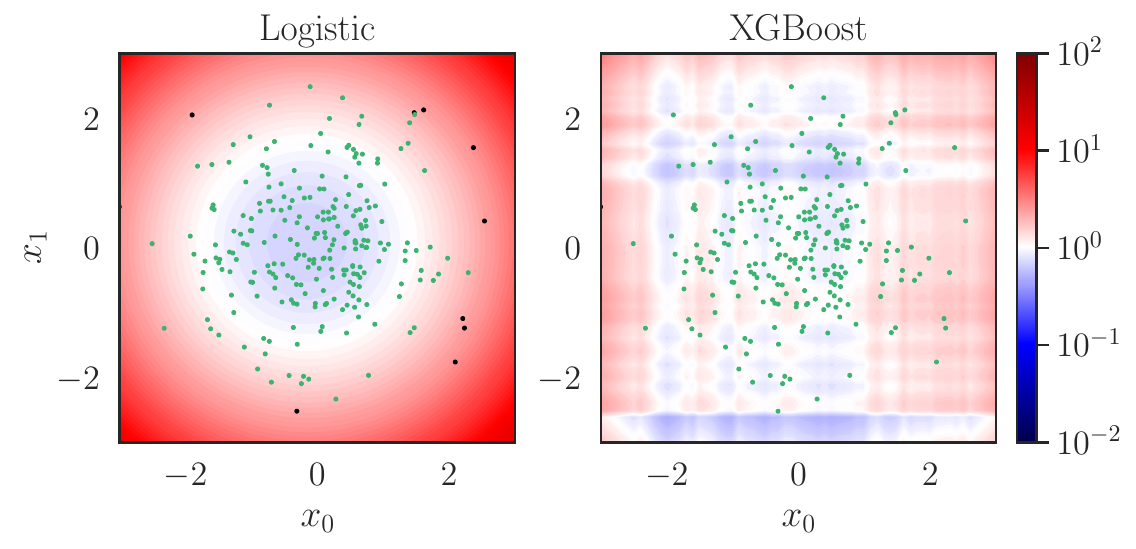}
        \caption{Degree of miscalibration for target population \texttt{C}.}
    \end{subfigure}
    \begin{subfigure}[b]{0.49\textwidth}
        \centering
        \includegraphics[width=0.975\linewidth]{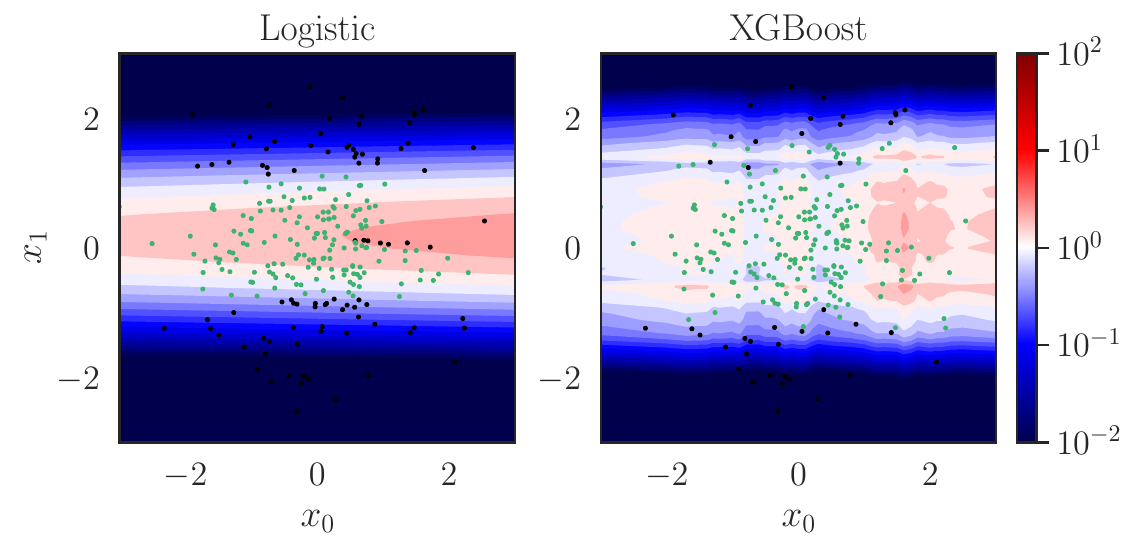}
        \caption{Degree of miscalibration for target population \texttt{D}.}    
    \end{subfigure}
    \caption{The actual degree of miscalibration for different target populations. The dots are a random subsample of the trial samples, where the green ones correspond to a degree of miscalibration within $\Gamma = 2$ and the black ones to a degree of miscalibration not bounded by $\Gamma = 2$.}
    \label{fig:degree_miscalibration:appx}
\end{figure}

For completeness, the actual degree of miscalibration for all target populations are visualized in Figure~\ref{fig:degree_miscalibration:appx} for the case without unmeasured selection factors $\Unobserved$.

\subsection{Seafood consumption policies}
\label{sec:nhanes_app}
The National Health and Nutrition Examination Survey data \citepalias{nhanes13} is produced by federal agencies and is in the public domain, allowing it to be reproduced without permission. In our evaluation the data is split into observational data $\setdata_0$ and trial data $\setdata_1$ based on the covariates age, income, gender and smoking history (age and income are standardized), e.g.
\begin{equation}
    \prob(\Sampling = 1|\X, \Unobserved) = 0.25 \cdot [f(\X_{\text{age}}) + f(\X_{\text{income}})] + 0.05 \cdot [\ind{\X_{\text{male}} = 1} + \ind{\X_{\text{smoking ever}} = 1}] + 0.3,
\end{equation}
where $f(\cdot)$ is the sigmoid function.


\end{document}